\documentclass[aps, pre, showpacs, superscriptaddress, twocolumn]{revtex4} 
\usepackage{graphicx}
\usepackage{dcolumn}
\usepackage{amsmath}
\usepackage{amssymb}
\usepackage{color}
\usepackage{bm}
\newcommand{\bvec}[1]{\mbox{\boldmath $#1$}}

\begin{document}
\preprint{HEP/123-qed}
\title{Nonequilibrium Kosterlitz-Thouless transition in a three-dimensional \\ driven disordered system}
\author{Taiki Haga}
\affiliation{Department of Physics, Kyoto University, Kyoto 606-8502, Japan}
\email[]{haga@scphys.kyoto-u.ac.jp}
\date{\today}

\begin{abstract}
We demonstrate a three-dimensional Kosterlitz-Thouless (KT) transition in the random field XY model driven out of thermal equilibrium.
By employing the spin-wave approximation and functional renormalization group approach, in the weak disorder regime, the three-dimensional driven random field XY model is found to exhibit a quasi-long-range order phase, wherein the correlation function shows power-law decay with a non-universal exponent that depends on the disorder strength.
This result is consistent with that reported in a previous numerical study.
We further develop a phenomenological theory of the three-dimensional KT transition by taking into account the effect of vortices.
The point of this theory is that the cross-section of the system with respect to a plane perpendicular to the driving direction is essentially identical to the two-dimensional pure XY model.
\end{abstract}

\pacs{11.10.Hi, 05.60.-k, 75.10.Nr}

\maketitle

\section{Introduction}
\label{sec:Introduction}

Two-dimensional (2D) systems with a global $U(1)$ symmetry such as liquid $^{4}{\rm He}$ films \cite{Bishop-78}, superconducting arrays of Josephson junctions \cite{Resnick-81}, and trapped atomic gases \cite{Hadzibabic-06} exhibit a “topologically ordered phase”, which is characterized by power-law decay of the correlation function with a continuously varying exponent.
The transition from such a quasi-long-range order (QLRO) phase to a disordered phase is called the Kosterlitz-Thouless (KT) transition \cite{Berezinskii-71,Kosterlitz-73,Nelson-77}.
The peculiarity of this transition comes from the fact that it is caused by the structural changes in topological defects or vortices.
More precisely, at low temperatures, vortices and antivortices form into bound pairs, and at some critical temperature, the dissociation of them occurs.
It is intriguing to understand the role of spatial dimensionality in the realization of the KT transition because the geometries and interactions of the topological defects crucially depend on the spatial dimensions.  
In the first step toward clarifying this problem, we ask whether there exists a topologically ordered phase and the KT transition in three dimensions.

Since a spatially inhomogeneous disorder can significantly change the long-distance physics of phase ordering systems, we investigate the possibility of higher-dimensional KT transition induced by the disorder.
For example, let us consider the random field XY model (RFXYM), where the two-component vector field is linearly coupled to a quenched random field.
The lower critical dimension of the RFXYM is known to be four \cite{Imry-75,Aizenman-89}.
Therefore, one may naively expect that the four-dimensional (4D) RFXYM exhibits the KT transition.
In fact, this model exhibits QLRO known as the Bragg glass phase below four dimensions \cite{Giamarchi-94,Feldman-00,Gingras-96,Menon-02}.
Although this is reminiscent of the topologically ordered phase in the 2D pure XY model, fundamental differences exist between them. 
First, in the Bragg glass phase of the RFXYM, the correlation function decays according to the power-law with a universal exponent and second, the transition to a disordered phase is considered as second order, not KT-like.
In addition, it is unclear whether the Bragg glass phase persists in three dimensions \cite{Fisher-97,LeDoussal-06,Tissier-06-NPRG,Tissier-06-2-loop}.
Therefore, at least in equilibrium, there is no example of a disorder-induced KT transition in three dimensions.

In the present paper, we show that the RFXYM exhibits a topologically ordered phase and the KT transition in three dimensions when it is driven at a uniform and steady velocity.
This result provides a simple example of a topological phase transition wherein the interplay between quenched disorder and nonequilibrium driving plays a crucial role.
From a simple argument, one finds that in the presence of the driving the lower critical dimension of the RFXYM becomes three, not four as in equilibrium.
Therefore, one may be led to predict that the driven random field XY model (DRFXYM) exhibits the KT transition in three dimensions.
The phase transition of this model was numerically investigated by us in Ref.~\cite{Haga-15}.
In this previous study, we calculated the correlation function for the nonequilibrium steady states and found that it exhibits power-law decay at weak disorder and low temperatures.
When the disorder strength and temperature increase, at some critical point we observed a transition to a disordered phase, wherein the correlation function decays exponentially.
We also determined the transition temperature as a function of the disorder strength and driving velocity by using the nonequilibrium relaxation method.
These numerical observations support the prediction that this model exhibits the three-dimensional (3D) KT transition.
However, numerical approach is insufficient to understand the large-scale behavior of the model because finite size effects are inevitable.
The purpose of this paper is to perform a renormalization group analysis of the DRFXYM and to show that it exhibits QLRO and the KT transition in three dimensions.

Let us recall the mechanism of the conventional KT transition in the 2D XY model before we outline the strategy of this study.
The QLRO at low temperatures is characterized by a line of fixed points.
This fixed line is a direct consequence of the fact that the spin-wave model of the 2D XY model is nothing but the massless free field theory.
In the QLRO phase, the only effect of the vortex-antivortex pairs is to renormalize the elastic constant (helicity modulus) of the spin-wave model.
At the transition point, the dissociation of them leads to the vanishing of the effective elastic constant and the QLRO is destroyed.
Therefore, the strategy of this study is as follows.
First, we show that the long-distance physics of the spin-wave model corresponding to the DRFXYM is essentially the same as that of the massless free field theory.
Second, we take into account the effect of the vortices by the renormalization of the elastic constant.

In the first part of this paper, we consider the spin-wave model in which the vortices are ignored.
It is an effective model that is valid only in the weak disorder regime.
By applying the functional renormalization group (FRG) theory, we show that this spin-wave model flows to the massless free field theory in the coarse-graining procedure, and that it exhibits QLRO, wherein the correlation function shows power-law decay with an exponent that depends on the disorder strength and the driving velocity.
We emphasize that this QLRO phase is quite different from the Bragg glass phase in the RFXYM, indeed the former is a consequence of a line of fixed points, while the latter is characterized by a single stable fixed point.
Therefore, the QLRO phase of the 3D-DRFXYM resembles the topologically ordered phase in the 2D XY model.

In the second part, we develop a phenomenological theory of the 3D KT transition by taking into account the effect of the vortices.
The elastic constant in the spin-wave model is renormalized due to the vortices.
To calculate this effective elastic constant, we invoke the so-called ``dimensional reduction'' property, which states that the long-distance physics of $D$-dimensional driven disordered systems is the same as that of $(D-1)$-dimensional pure systems.
With the aid of this property, we derive the flow equation of the effective elastic constant, which is similar to that for the 2D pure XY model, except that the temperature is replaced with the disorder strength.
The dimensional reduction also enables us to discuss the changes in the vortex structure at the transition point. 
The vortices in the 3D XY model are lines, not points in contrast to the 2D case.
The dissociation of the vortex-antivortex pairs in the 2D XY model corresponds to the breakdown of vortex rings in the 3D-DRFXYM.

This paper is organized as follows.
In Sec.~\ref{Sec:Model}, we define the DRFXYM and its spin-wave model.
We show that the lower critical dimension of the DRFXYM is three.
In Sec.~\ref{Sec:Dimensional_reduction}, we derive a dimensional reduction property for the DRFXYM from an intuitive argument.
It predicts that the large-scale behavior of the 3D-DRFXYM at zero temperature is identical to that of the 2D pure XY model.
We also emphasize that this dimensional reduction does not always hold.
In Sec.~\ref{Sec:RG_analysis_of_the_spin-wave_model}, we perform the FRG analysis of the spin-wave model.
We show that the FRG equation for the second cumulant of the disorder has a line of fixed point, as in the 2D XY model.
In Sec.~\ref{Sec:Effects_of_the_vortices}, we consider the effects of the vortices, which are ignored in the spin-wave model.
The effective elastic constant (helicity modulus) is calculated with the aid of the dimensional reduction property.
The changes in the vortex structure at the transition point are also discussed.

\section{Model}
\label{Sec:Model}

Let $\bvec{\phi}(\bvec{r})=(\phi^1(\bvec{r}),\phi^2(\bvec{r}))$ be a two component real vector field.
The Hamiltonian of the $D$-dimensional XY model with a quenched random field $\bvec{h}(\bvec{r})=(h^1(\bvec{r}),h^2(\bvec{r}))$ is given by
\begin{eqnarray}
H[\bvec{\phi};\bvec{h}]=\int {\rm d}^D \bvec{r} \biggl[ \frac{1}{2} K |\bvec{\nabla} \bvec{\phi}|^2+U(\rho) - \bvec{h} \cdot \bvec{\phi} \biggr],
\label{Hamiltonian}
\end{eqnarray}
where $\rho=|\bvec{\phi}|^2/2$ is the field amplitude and $U(\rho)=(g/2)(\rho-1/2)^2$ is a double-well potential.
The random field $h^{\alpha}(\bvec{r})$ obeys a mean-zero Gaussian distribution with 
\begin{equation}
\overline{ h^{\alpha}(\bvec{r}) h^{\beta}(\bvec{r'}) } =h_0^2 \delta^{\alpha \beta} \delta(\bvec{r}-\bvec{r'}),
\end{equation}
where the over-bar represents the average over the quenched disorder.
The dynamics of the field $\bvec{\phi}(\bvec{r},t)$ is described by 
\begin{equation}
\partial_t \phi^{\alpha}+v \partial_x \phi^{\alpha} =- \frac{\delta H[\bvec{\phi};\bvec{h}]}{\delta \phi^{\alpha}}+\xi^{\alpha},
\label{EM}
\end{equation}
where $v$ is a uniform and time-independent driving velocity, and $\xi^{\alpha}(\bvec{r},t)$ represents the thermal noise that satisfies 
\begin{equation}
\langle \xi^{\alpha}(\bvec{r},t) \xi^{\beta}(\bvec{r'},t') \rangle =2 T \delta^{\alpha \beta} \delta(\bvec{r}-\bvec{r'}) \delta(t-t').
\end{equation}
We call this model the driven random field XY model (DRFXYM).
It describes the relaxation dynamics of phase ordering systems driven in a random environment.
Examples of such systems include liquid crystals flowing in porous media \cite{Araki-12}.
The irregular surface structure of the porous substrate results in symmetry breaking random anchoring, which is similar to the random field in the XY model.

The original model defined by Eqs.~(\ref{Hamiltonian}) and (\ref{EM}) is too complicated for renormalization group (RG) analysis.
If one is interested in the phase structure at weak disorder, it is convenient to introduce the spin-wave model of the DRFXYM.
We ignore the amplitude fluctuation of $\bvec{\phi}$ and define the single-valued phase parameter $u \in (-\infty, \infty)$ by $(\phi^1,\phi^2) = (\cos u , \sin u)$.
From Eq.~(\ref{EM}), the dynamics of $u(\bvec{r},t)$ is described by
\begin{eqnarray}
\partial_t u + v \partial_x u = K \nabla^2 u + F(\bvec{r};u) + \xi(\bvec{r},t),
\label{EM_SW}
\end{eqnarray}
where $F(\bvec{r};u) = -h^1(\bvec{r}) \sin u + h^2(\bvec{r}) \cos u$ is a random force. 
Its second cumulant is written as
\begin{equation}
\overline{ F(\bvec{r};u) F(\bvec{r'};u') } = \Delta_{\mathrm{B}}(u-u') \delta(\bvec{r}-\bvec{r'}),
\label{bare_cumulant}
\end{equation}
with $\Delta_{\mathrm{B}}(u) = h_0^2 \cos u$, where the subscript ``$\mathrm{B}$'' represents the ``bare'' random force.
The thermal noise $\xi(\bvec{r},t)$ satisfies
\begin{equation}
\langle \xi(\bvec{r},t) \xi(\bvec{r'},t') \rangle = 2 T \delta(\bvec{r}-\bvec{r'}) \delta(t-t').
\end{equation}
This model was also introduced in the context of the moving Bragg glass in Refs.~\cite{Giamarchi-96,LeDoussal-98,Balents-98} to describe the dynamics of the displacement field of an elastic lattice driven in a random pinning potential.
The spin-wave model is valid only when the order parameter varies slowly in space.
Thus, it is not reliable in the strong disorder regime.

Since we consider the nonequilibrium steady states of this model, in the following, $\langle...\rangle$ denotes the average over the distribution function of the steady state,
\begin{equation}
\langle A[\bvec{\phi}] \rangle \equiv \int \mathcal{D} \bvec{\phi} A[\bvec{\phi}] P_{\mathrm{st}}[\bvec{\phi};\bvec{h}],
\end{equation}
where $P_{\mathrm{st}}$ is the probability distribution function of the steady state for a given realization of the random field. 
The disorder average is given by
\begin{equation}
\overline{ \langle A[\bvec{\phi}] \rangle } \equiv \int \mathcal{D} \bvec{h} \langle A[\bvec{\phi}] \rangle P_{\mathrm{R}}[\bvec{h}],
\end{equation}
where $P_{\mathrm{R}}$ is the distribution function of the random field.

Let us consider the lower critical dimension of the DRFXYM.
At zero temperature, the stationary state of the spin-wave model satisfies
\begin{eqnarray}
-K \nabla^2 u + v \partial_x u = F(\bvec{r};u).
\end{eqnarray}
If we ignore the field dependence of $F(\bvec{r};u)$, the correlation function behaves as
\begin{equation}
\overline{ \langle u(\mathbf{q}) u(-\mathbf{q}) \rangle } \sim \frac{h_0^2}{K^2 |\mathbf{q}|^4+v^2 q_x^2},
\label{Spin_wave_correlation}
\end{equation}
whose $\mathbf{q}$-integral exhibits an infrared-divergence below three dimensions, thus the lower critical dimension is three.
From the analogy to the 2D pure XY model, the 3D-DRFXYM is expected to exhibit QLRO in the weak disorder regime.
In fact, Eq.~(\ref{Spin_wave_correlation}) leads to the power-law decay of the correlation function, 
\begin{equation}
C(\bvec{r}) = \overline{ \langle \bvec{\phi}(\bvec{r}) \cdot \bvec{\phi}(\bvec{0}) \rangle } = \overline{ \langle e^{i(u(r)-u(0))} \rangle } \sim |\bvec{r}|^{-\eta},
\label{correlation_power_law}
\end{equation}
which is anisotropic due to the driving.
The exponents here are given by $\eta_{\parallel}= h_0^2/(8 \pi Kv)$ for the direction parallel to the driving velocity and $\eta_{\perp}= h_0^2/(4 \pi Kv)$ for the perpendicular direction \cite{Haga-15}.
These exponents reasonably agree with those obtained from numerical simulations in Ref.~\cite{Haga-15}.

It is worth to note that for the equilibrium case ($v=0$), Eq.~(\ref{Spin_wave_correlation}) also predicts the power-law decay of $C(\bvec{r})$ with an exponent $\eta=h_0^2/(8 \pi^2 K^2)$ at $D=4$.
Therefore, one may expect that the 4D-RFXYM exhibits the KT transition.
However, as mentioned in Sec.~\ref{sec:Introduction}, this is incorrect (See Sec.~\ref{sec:RG_evolution_Equilibrium_case}).

\section{Dimensional reduction}
\label{Sec:Dimensional_reduction}

In equilibrium, standard perturbation theory predicts that the critical exponents of $D$-dimensional random field spin models are the same as those of $(D-2)$-dimensional pure spin models \cite{Aharony-76,Parisi-79}.
This remarkable property is called the dimensional reduction.
However, this dimensional reduction breaks down in low enough dimensions.
For example, the lower critical dimension of the random field Ising model (RFIM) is known to be two from phenomenological and rigorous arguments \cite{Imry-75,Aizenman-89,Bricmont-87}.
On the other hand, the dimensional reduction predicts that it is three because the lower critical dimension of the pure Ising model is one.
It is well-known that the cause of this failure is the presence of multiple local minima in the energy landscape \cite{Nattermann-97}.

We now derive a novel type of dimensional reduction property for driven disordered systems.
It predicts that the critical behavior of the $D$-dimensional DRFXYM at zero temperature is identical to that of the $(D-1)$-dimensional pure XY model in equilibrium.
This property will play a crucial role when we discuss the effect of the vortices in Sec.~\ref{Sec:Effects_of_the_vortices}.
At zero temperature, Eq.~(\ref{EM}) is written as
\begin{eqnarray}
\partial_t \bvec{\phi}+v \partial_x \bvec{\phi} = K \nabla^2 \bvec{\phi} - U'(\rho) \bvec{\phi} + \bvec{h}(\bvec{r}).
\label{EM_zero_temp}
\end{eqnarray}
After a sufficiently long time, the solution of Eq.~(\ref{EM_zero_temp}) reaches a stationary state $\bvec{\phi}_{\mathrm{st}}(\bvec{r})$, which satisfies the following equation:
\begin{eqnarray}
v \partial_x \bvec{\phi}_{\mathrm{st}} = K \nabla^2 \bvec{\phi}_{\mathrm{st}} - U'(\rho_{\mathrm{st}}) \bvec{\phi}_{\mathrm{st}} + \bvec{h}(\bvec{r}),
\label{EM_st}
\end{eqnarray}
where $\rho_{\mathrm{st}}=|\bvec{\phi}_{\mathrm{st}}|^2/2$.
We assume that there is only a single stationary state.
In the large length scale, the longitudinal elastic term $K \partial_x^2 \bvec{\phi}$ is negligible compared to the advection term $v \partial_x \bvec{\phi}$.
Thus, Eq.~(\ref{EM_st}) becomes
\begin{eqnarray}
v \partial_x \bvec{\phi} = K \nabla_{\perp}^2 \bvec{\phi} - U'(\rho) \bvec{\phi} + \bvec{h}(x,\bvec{r}_{\perp}),
\label{EM_st_perp}
\end{eqnarray}
where $\nabla_{\perp}$ is the derivative operator for the transverse directions and $\bvec{r}_{\perp}$ represents the transverse coordinate.
If the coordinate $x$ is considered to be a fictitious time and $\bvec{h}(x,\bvec{r}_{\perp})$ as thermal noise, Eq.~(\ref{EM_st_perp}) is nothing but the dynamical equation for the $(D-1)$-dimensional pure XY model with a temperature 
\begin{equation}
T_{\mathrm{eff}} = \frac{h_0^2}{2v}.
\label{effective_temperature}
\end{equation}
Eq.~(\ref{EM_st_perp}) has infinitely many solutions because one can obtain one of them by solving this equation along $x$-direction starting from an arbitrary ``initial condition'' $\bvec{\phi}(x=0,\bvec{r}_{\perp})$.
However, there exist a solution $\bvec{\phi}_{*}(x,\bvec{r}_{\perp})$ of Eq.~(\ref{EM_st_perp}) such that its large-scale behavior is the same as that of $\bvec{\phi}_{\mathrm{st}}(\bvec{r})$.
Therefore, one may naively expect that the transverse section of the $D$-dimensional DRFXYM at zero temperature is identical to the $(D-1)$-dimensional pure XY model.
Recall that, in the 2D XY model, the correlation function shows power-law decay, $C(\bvec{r}) \sim |\bvec{r}|^{-\eta_{\mathrm{2D}}}$ with $\eta_{\mathrm{2D}}=T/(2\pi K)$ at low temperatures.
The dimensional reduction implies that, in the 3D-DRFXYM, the correlation function for the transverse direction $(\bvec{r} \perp \bvec{e}_x)$ also shows power-law decay, $C(\bvec{r}) \sim |\bvec{r}|^{-\eta_{\perp}}$, where the exponent $\eta_{\perp}$ can be obtained by replacing the temperature $T$ in $\eta_{\mathrm{2D}}$ with the effective temperature Eq.~(\ref{effective_temperature}),
\begin{equation}
\eta_{\perp} = \frac{T_{\mathrm{eff}}}{2\pi K} = \frac{h_0^2}{4\pi Kv}.
\label{eta_SW}
\end{equation}
This value agrees with that obtained from the simple spin-wave argument, Eqs.~(\ref{Spin_wave_correlation}) and (\ref{correlation_power_law}).

However, this dimensional reduction is not always correct.
We show a simple counterexample.
Let us consider the driven random field Ising model (DRFIM), which is defined by Eqs.~(\ref{Hamiltonian}) and (\ref{EM}) with a one-component scalar field $\phi(\bvec{r})$.
In equilibrium, the lower critical dimension of the RFIM is two as mentioned above.
Since the advection term $v\partial_x \phi$ reduces the lower critical dimension, we expect that the 2D-DRFIM exhibits long-range order at weak disorder.
However, the dimensional reduction predicts that it does not because it is identical to the one-dimensional pure Ising model.

The breakdown of the dimensional reduction is a consequence of the fact that there are a large number of stationary states satisfying Eq.~(\ref{EM_st}), and that they contribute to physical quantities, such as the correlation function, with a non-trivial weight.
This situation is analogous to that of the conventional dimensional reduction in equilibrium.
The remarkable difference from the equilibrium cases is that the concept of the energy landscape is meaningless because the advection term $v\partial_x \bvec{\phi}$ in Eq.~(\ref{EM}) cannot be cast into the functional derivative of an appropriate potential.

\section{FRG analysis of the spin-wave model}
\label{Sec:RG_analysis_of_the_spin-wave_model}

The standard perturbative approach leads to the dimensional reduction, which is found to be incorrect.
To overcome this difficulty, the functional renormalization group (FRG) theory has been developed for disordered systems in equilibrium \cite{Giamarchi-94,Feldman-00,Fisher-85,Fisher-86,Balents-93,Feldman-02,LeDoussal-03,LeDoussal-04}.
In this formalism, one constructs the RG flow of a whole function of the disorder cumulant $\Delta(u)$ (See Eq.~(\ref{bare_cumulant})).
At a fixed point corresponding to this RG flow, the renormalized cumulant can have a linear cusp as a function of the field, $\Delta(u) \simeq \Delta(0)+\Delta'(0^+)|u|$.
Such non-analytic behavior is a consequence of the presence of multiple stationary states and leads to the breakdown of the dimensional reduction.
We expect that this relation between the analyticity of the disorder cumulant and the dimensional reduction also holds for the nonequilibrium cases.
In the following, we perform the FRG analysis of the spin-wave model (\ref{EM_SW}) and show that it flows to the massless free field theory in the large-scale limit. 
This means that the dimensional reduction holds within the spin-wave approximation.

To derive the flow equation of the renormalized disorder cumulant, we employ the non-perturbative FRG (NP-FRG) approach developed in Refs.~\cite{Tarjus-08} and \cite{Tissier-12}.
This formalism is a hybrid of the FRG and the so-called non-perturbative RG approach \cite{Wetterich-02,Canet-05,Canet-10,Canet-11}.
It enables us to go beyond the leading order flow equation more systematically compared to the perturbative FRG approach.

\subsection{Exact flow equation for the effective action}

We first recall how the equation of motion (\ref{EM_SW}) can be cast into a field theoretical formalism.
By introducing the replicated fields $U_a = {}^t( u_a, \hat{u}_a ), \: a=1,...,n$, the disorder averaged action is given by
\begin{eqnarray}
S[\{ U_a \}] &=& \sum_a \int_{rt} \hat{u}_a \bigl[ \partial_t u_a - T \hat{u}_a + v \partial_x u_a - K \nabla^2 u_a \bigr] \nonumber \\
&&-\frac{1}{2} \sum_{a,b} \int_{rtt'} \hat{u}_{a,t} \hat{u}_{b,t'} \Delta_{\mathrm{B}}( u_{a,t} - u_{b,t'} ),
\label{Bare_action}
\end{eqnarray}
where $\Delta_{\mathrm{B}}(u) = h_0^2 \cos u$ is the second cumulant of the bare random force.
The detailed derivation of Eq.~(\ref{Bare_action}) is presented in Appendix \ref{Appendix:Derivation_of_the_dynamical_action}.
We next introduce source fields $J_a={}^t(j_a,\hat{j}_a), \: a=1,...,n$, and define  the generating functional $W[\{J_a\}]$ by,
\begin{eqnarray}
e^{W[\{J_a\}]} &=& \int \mathcal{D}U \exp \biggl[ -S[\{ U_a \}] \nonumber \\
&&+ \sum_a \int_{rt} {}^t J_a \cdot U_a  \biggr].
\end{eqnarray}
The effective action is defined as a Legendre transform:
\begin{eqnarray}
\Gamma[\{ U_a \}] = - W[\{J_a\}] + \sum_a \int_{rt} {}^t J_a \cdot U_a,
\label{Full_effective_action}
\end{eqnarray}
where $U_a$ and $J_a$ are related by 
\begin{equation}
U_a = \frac{\delta W[\{J_a\}]}{\delta J_a}.
\end{equation}
Since $\Gamma[\{ U_a \}]$ gives the renormalized vertices, its zero momentum limit defines the renormalized disorder.

The NP-FRG formalism is based on an exact flow equation for the scale-dependent effective action $\Gamma_k[\{ U_a \}]$, which includes only high energy modes with momenta larger than the running scale $k$.
As $k$ goes from the cutoff scale $\Lambda$ to zero, $\Gamma_k$ interpolates between the bare action Eq.~(\ref{Bare_action}) and the full effective action Eq.~(\ref{Full_effective_action}).
To define the scale-dependent effective action $\Gamma_k$, we add to the original action a momentum-dependent mass term
\begin{equation}
\Delta S_k[\{ U_a \}] = \frac{1}{2} \sum_{a} \int_{q} {}^t U_a(q) \: \mathbf{R}_k(\mathbf{q}) \: U_a(-q),
\end{equation}
where $q=(\mathbf{q},\omega)$ and $\int_q=\int d^{D} \mathbf{q} d \omega/(2\pi)^{D+1}$.
A frequency independent matrix $\mathbf{R}_k(\mathbf{q})$ is given by
\begin{eqnarray}
\bvec{\mathrm{R}}_k(\mathbf{q}) = \left(
\begin{array}{ccc}
0 & R_k(\mathbf{q}) \\
R_k(\mathbf{q}) & 0
\end{array}
\right),
\label{R_matrix}
\end{eqnarray}
where $R_k(\mathbf{q})$ is an infrared cutoff function, which has a constant value proportional to $k^2$ for $|\mathbf{q}| \ll k$ and rapidly decreases for $|\mathbf{q}| > k$.
The explicit form of $R_k(\mathbf{q})$ will be specified later.
Note that the off-diagonal form of Eq.~(\ref{R_matrix}) leads to the term $R_k(\mathbf{q}) \hat{u}_a(q) u_a(-q)$, which suppresses the fluctuations with momenta smaller than $k$.
The scale-dependent generating functional $W_k[\{J_a\}]$ is defined by
\begin{eqnarray}
e^{W_k[\{J_a\}]} &=& \int \mathcal{D}U \exp \biggl[ -S[\{ U_a \}] - \Delta S_k[\{ U_a \}] \nonumber \\
&&+ \sum_a \int_{rt} {}^t J_a \cdot U_a  \biggr].
\end{eqnarray}
Then, the scale-dependent effective action is given by
\begin{eqnarray}
\Gamma_k[\{ U_a \}] &=& -W_k[\{ J_a \}] + \sum_a \int_{rt} {}^t J_a \cdot U_a \nonumber \\
&&- \Delta S_k[\{ U_a \}],
\end{eqnarray}
where $U_a$ and $J_a$ are related by 
\begin{equation}
U_a = \frac{\delta W_k[\{J_a\}]}{\delta J_a}.
\end{equation}
It can be shown that $\Gamma_{k=0}=\Gamma$ and $\lim_{k \to \infty} \Gamma_k = S$.

The flow of $\Gamma_k$ is described by the Wetterich equation \cite{Wetterich-02},
\begin{equation}
\partial_k \Gamma_k = \frac{1}{2} \mathrm{Tr} \partial_k  \hat{\mathbf{R}}_k(\mathbf{q}) \Bigl[ \Gamma_k^{(2)} + \hat{\mathbf{R}}_k(\mathbf{q}) \Bigr]^{-1},
\label{exact_flow_equation}
\end{equation}
where $\Gamma_k^{(2)}$ is the second functional derivative of $\Gamma_k$ and ``$\mathrm{Tr}$'' represents an integration over momentum and frequency as well as a sum over replica indices and the two conjugate fields $\{ u, \hat{u} \}$.
We have introduced a $2n \times 2n$ matrix
\begin{eqnarray}
\hat{\bvec{\mathrm{R}}}_k(\mathbf{q}) = \bvec{\mathrm{R}}_k(\mathbf{q}) \otimes \mathbf{I}_n,
\end{eqnarray}
where $\mathbf{I}_n$ is the $n \times n$ unit matrix, which acts on the space of the replica index.

According to Ref.~\cite{Tarjus-08}, $\Gamma_k$ is expanded by increasing the number of free replica sums as
\begin{equation}
\Gamma_k[\{ U_a \}] = \sum_{p=1}^{\infty} \sum_{a_1,...,a_p} \frac{(-1)^{p-1}}{p!} \Gamma_{p,k}[U_{a_1},...,U_{a_p}],
\label{Gamma_expansion}
\end{equation}
where $\Gamma_{p,k}$ corresponds to the $p$-th cumulant of the renormalized disorder.
Insertion of Eq.~(\ref{Gamma_expansion}) into Eq.~(\ref{exact_flow_equation}) leads to the exact flow equations for $\Gamma_{p,k}$.
In Appendix \ref{Appendix:Exact_flow_equations}, the exact flow equations for $\Gamma_{1,k}$, $\Gamma_{2,k}$, and $\Gamma_{3,k}$ are given by Eqs,~(\ref{Appendix_gamma_1}), (\ref{Appendix_gamma_2}), and (\ref{Appendix_gamma_3}), respectively.
The exact flow equation for $\Gamma_{p,k}$ contains $\Gamma_{p+1,k}$, thus we have an infinite hierarchy of the coupled flow equations.

\subsection{Flow equations for the disorder cumulants}

To solve the exact flow equations, we have to introduce approximations for the functional forms of $\Gamma_{p,k}$.
We employ the following ansatz for the one-replica part,
\begin{equation}
\Gamma_{1,k} = \int_{rt} \hat{u} \bigl[ X_k (\partial_t u - T_k \hat{u}) + v \partial_x u - K \nabla^2 u \bigr],
\label{Gamma_1}
\end{equation}
where $X_k$ and $T_k$ are the scale-dependent relaxation coefficient and temperature. For the multi-replica part,
\begin{equation}
\Gamma_{p,k} = \int_{rt_1...t_p} \hat{u}_{a_1,t_1} ... \hat{u}_{a_p,t_p} \Delta_{p,k}( u_{a_1,t_1}, ... , u_{a_p,t_p} ),
\label{Gamma_n}
\end{equation} 
where $\Delta_{p,k}(u_1,...,u_p)$ is the $p$-th cumulant of the renormalized random force, which is a fully symmetric function satisfying
\begin{equation}
\Delta_{p,k}(u_1+2\pi,u_2,...,u_p) = \Delta_{p,k}(u_1,u_2,...,u_p),
\end{equation}
\begin{equation}
\Delta_{p,k}(u_1+\lambda,...,u_p+\lambda) = \Delta_{p,k}(u_1,...,u_p),
\end{equation} 
for an arbitrary $\lambda$.
Although the bare random force is chosen as Gaussian, the higher-order cumulants can be generated in the coarse-graining procedure.
Note that the elastic constant $K$ and the driving velocity $v$ in Eq.~(\ref{Gamma_1}) are not renormalized \cite{LeDoussal-98}, while the relaxation coefficient $X_k$ and temperature $T_k$ can be renormalized.
From the functional form Eq.~(\ref{Gamma_n}), the RG equation for $\Delta_p$ is obtained from
\begin{equation}
\partial_k \Delta_{p,k}(u_1,...,u_p) = \frac{\delta^p}{\delta \hat{u}_1... \delta \hat{u}_p } \partial_k \Gamma_{p,k}[U_1,...,U_p],
\label{RGeq_Delta-0}
\end{equation}
where the functional derivative is evaluated for a uniform field configuration: $ u_{1,rt} \equiv u_1,...,u_{p,rt} \equiv u_p $ and  $ \hat{u}_{1,rt} \equiv 0,.., \hat{u}_{p,rt} \equiv 0 $.

From Eqs.~(\ref{Appendix_gamma_2}), (\ref{Appendix_gamma_3}), (\ref{Gamma_1}), (\ref{Gamma_n}), and (\ref{RGeq_Delta-0}), we obtain the flow equation for $\Delta_{p,k}$.
The exact flow equations for $\Gamma_{p,k}$ contain the one-replica propagator,
\begin{equation}
\mathrm{P}[U] = \bigl[ \Gamma_{1,k}^{(2)}[U] + \mathbf{R}_k \bigr]^{-1},
\end{equation}
whose matrix elements are written as
\begin{eqnarray}
P_{11}(\mathbf{q},\omega) &=& \frac{2 X_k T_k}{D(\mathbf{q},\omega)}, \nonumber \\
P_{12}(\mathbf{q},\omega) &=& \frac{M(\mathbf{q})-i(X_k \omega - q_x v)}{D(\mathbf{q},\omega)}, \nonumber \\
P_{21}(\mathbf{q},\omega) &=& \frac{M(\mathbf{q})+i(X_k \omega - q_x v)}{D(\mathbf{q},\omega)}, \nonumber \\
P_{22}(\mathbf{q},\omega) &=& 0,
\end{eqnarray}
where 
\begin{eqnarray}
M(\mathbf{q}) &=& K|\mathbf{q}|^2 + R_k(\mathbf{q}), \nonumber \\
D(\mathbf{q},\omega) &=& M(\mathbf{q})^2 +(X_k \omega - q_x v)^2.
\end{eqnarray}
We will use notations such as $P_{12}(\mathbf{q})=P_{12}(\mathbf{q},\omega=0)$ and $D(\mathbf{q})=D(\mathbf{q},\omega=0)$.
Below, instead of $k$, we use a renormalization scale $l=-\ln (k/\Lambda)$, which moves from $0$ to $\infty$ as $k$ goes from $\Lambda$ to $0$.
To express the flow equation in a compact form, we also define the following integrals,
\begin{eqnarray}
L_n^- &=& -\frac{1}{2} \int_{\mathbf{q}} \partial_l R_k(\mathbf{q}) \bigl\{ n P_{21}(\mathbf{q})^{n+1} + n P_{12}(\mathbf{q})^{n+1} \bigr\}, \nonumber \\
L_n^+ &=& -\frac{1}{2} \int_{\mathbf{q}} \partial_l R_k(\mathbf{q}) \sum_{j=1}^n  2 P_{21}(\mathbf{q})^{n+1-j} P_{12}(\mathbf{q})^j,
\label{def_L}
\end{eqnarray}
where $\partial_l=-k\partial_k$.

We consider the zero temperature case $T=T_k=0$. 
From Eq.~(\ref{Appendix_gamma_2}), the flow equation for $\Delta_2$ is given as follows:
\begin{eqnarray}
\partial_l \Delta_2(u_1,u_2) = \frac{1}{2} \Bigl[ (1) + (2) + (3) + (4) + \mathrm{perm} \Bigr],
\label{RG_Delta2_dimensionfull}
\end{eqnarray}
\begin{eqnarray}
(1) = \Delta_2^{(11)}(u_1,u_2) \Delta_2(u_1,u_2) L_2^+, \nonumber
\end{eqnarray}
\begin{eqnarray}
(2) = \Delta_2^{(10)}(u_1,u_2) \Delta_2^{(01)}(u_1,u_2) L_2^-, \nonumber
\end{eqnarray}
\begin{eqnarray}
(3) = \Delta_2^{(20)}(u_1,u_2) \Delta_2(u_1,u_1) L_2^+, \nonumber
\end{eqnarray}
\begin{eqnarray}
(4) = - 2 \Delta_3^{(100)}(u_1,u_1,u_2) L_1^-, \nonumber
\end{eqnarray}
where ``$\mathrm{perm}$'' denotes the expression obtained by permuting $u_1$ and $u_2$, and we have used simplified notations such as 
\begin{eqnarray}
\Delta_2^{(11)}(u_a,u_b) &=& \partial_{u_1} \partial_{u_2} \Delta_2(u_a,u_b), \nonumber \\
\Delta_2^{(20)}(u_a,u_b) &=& \partial_{u_1} \partial_{u_1} \Delta_2(u_a,u_b).
\end{eqnarray}
From Eq.~(\ref{Appendix_gamma_3}), one can also obtain the flow equation for $\Delta_3$, which contains terms proportional to $\Delta_2^3$, $\Delta_2 \Delta_3$, and $\Delta_4$.
Since it is rather complicated, we will present in Appendix \ref{Appendix:Flow_equation_for_Delta_3}.

\subsubsection{Equilibrium case}

We first consider the equilibrium case ($v=0$).
It is convenient that the momentum $\mathbf{q}$ is measured in units of the running scale $k$,
\begin{equation}
y= \frac{|\mathbf{q}|^2}{k^2}.
\end{equation}
The cutoff function $R_k(\mathbf{q})$ is written as
\begin{eqnarray}
R_k(\mathbf{q}) = K k^2 r(y).
\label{cutoff_function_eq}
\end{eqnarray}
For simplicity, we employ the ``optimized'' cutoff function \cite{Litim-00},
\begin{eqnarray}
r(y) = (1-y) \Theta(1-y),
\label{optimized_cutoff_function}
\end{eqnarray}
where $\Theta(x)$ is the step function.
By using this cutoff function, the integrals in Eq.~(\ref{def_L}) are calculated as
\begin{eqnarray}
L_n^- = L_n^+ = 2n K^{-n} k^{D-2n} \frac{4}{D} A_D,
\end{eqnarray}
where ${A_D}^{-1}=2^{D+1} \pi^{D/2} \Gamma(D/2)$.
We rewrite Eq.~(\ref{RG_Delta2_dimensionfull}) in a scaled form by introducing renormalized dimensionless quantities.
In the following, the cutoff scale $\Lambda$ is set to unity.

The dimensionless cumulants are defined by
\begin{equation}
\delta_{2}(u_1,u_2) = \frac{16}{D} A_D K^{-2} k^{D-4} \Delta_{2}(u_1,u_2),
\label{def_delta_eq}
\end{equation}
\begin{equation}
\delta_{3}(u_1,u_2,u_3) = \left(\frac{16}{D}\right)^2 {A_D}^2 K^{-3} k^{2D-6} \Delta_{3}(u_1,u_2,u_3).
\label{def_delta3_eq}
\end{equation}
We also define the following notation:
\begin{equation}
\delta(u_a-u_b) = \delta_2(u_a,u_b), 
\label{def_delta}
\end{equation}
\begin{equation}
\delta_3'(u_a-u_b) = \frac{1}{2} \bigl\{ \delta_3^{(100)}(u_a,u_a,u_b) + \delta_3^{(100)}(u_b,u_b,u_a) \bigr\}.
\label{def_delta3}
\end{equation}
The RG equation for $\delta(u)$ is given by
\begin{eqnarray}
\partial_l \delta(u) &=& - (D-4) \delta(u) + \delta''(u) (\delta(0)-\delta(u)) \nonumber \\
&&- \delta'(u)^2 - \delta_3'(u).
\label{RGeq_delta_eq}
\end{eqnarray}
Eq.~(\ref{RGeq_delta_eq}) without the third order cumulant $\delta_3'(u)$ is first derived by Fisher in Ref.~\cite{Fisher-85} at weak disorder.
The RG equation for $\delta_3'(u)$ is presented in Appendix \ref{Appendix:Flow_equation_for_Delta_3}.

The long-distance physics of the system is controlled by a fixed point of Eq.~(\ref{RGeq_delta_eq}).
In fact, at a critical point, the correlation function decays as $C(\bvec{r}) = \overline{ \langle e^{i(u(r)-u(0))} \rangle } \sim |\bvec{r}|^{-\eta}$ with
\begin{equation}
\eta = \delta_*(0),
\label{eta_delta_eq}
\end{equation}
where $\delta_*(0)$ is the fixed point value of $\delta(0)$.
The derivation of Eq.~(\ref{eta_delta_eq}) is given in Appendix \ref{Appendix:Relation_between_exponent and_disorder}.

\subsubsection{Nonequilibrium case}
\label{sec:flow_equations_neq}

For the nonequilibrium case, it is convenient that the transverse momentum $\mathbf{q}_{\perp}$ and longitudinal momentum $q_x$ are measured in units of $k$ and $k^2$,  respectively, considering the anisotropy of the system due to the driving,
\begin{equation}
y_{\perp} = \frac{|\mathbf{q}_{\perp}|^2}{k^2}, \:\:\: y_{\parallel} = \frac{q_x^2}{k^4}.
\end{equation}
We employ an infrared cutoff function independent of $q_x$, 
\begin{equation}
R_k(\mathbf{q})=K k^2 r (y_{\perp}) = K k^2 (1-y_{\perp}) \Theta(1-y_{\perp}).
\label{cutoff_function_neq}
\end{equation}
The integrals in Eq.~(\ref{def_L}) are calculated as follows:
\begin{eqnarray}
L_n^{\pm} = \frac{4}{D-1}A_{D-1} K^{-n+1} k^{D-2n+1} v^{-1} l_n^{\pm}(z_k),
\end{eqnarray}
where
\begin{equation} 
z_k = v^{-2} K^2 k^2 = v^{-2} K^2 e^{-2l},
\label{def_z}
\end{equation}
which is related to the ratio of the longitudinal elastic term $K \partial_x^2 u$ to the advection term $v \partial_x u$, and
\begin{eqnarray}
l_n^-(z) &=& \frac{n}{\pi} \int_{-\infty}^{\infty} dx \: (1+z x^2 + i x)^{-(n+1)},
\label{def_l_n-}
\end{eqnarray}
\begin{eqnarray}
l_n^+(z) &=& \frac{1}{\pi} \int_{-\infty}^{\infty} dx \: \sum_{j=1}^n (1+z x^2 + i x)^{-(n+1-j)} \nonumber \\
&&\times (1+z x^2 - i x)^{-j}.
\label{def_l_n+}
\end{eqnarray}
One can easily check that $l_n^+(0)=1$ while $l_n^-(z) \sim z^n $ for a small $z$.

The dimensionless cumulants are defined by 
\begin{eqnarray}
\delta_{2}(u_1,u_2) &=& \frac{4}{D-1}A_{D-1} K^{-1} v^{-1} k^{D-3} \nonumber \\
&&\times \Delta_{2}(u_1,u_2), 
\label{def_delta_neq}
\end{eqnarray}
\begin{eqnarray}
\delta_{3}(u_1,u_2,u_3) &=& \left(\frac{4}{D-1}\right)^2 {A_{D-1}}^2 K^{-1} v^{-2} k^{2D-4} \nonumber \\
&&\times \Delta_{3}(u_1,u_2,u_3),
\label{def_delta3_neq}
\end{eqnarray}
and $\delta(u)$ and $\delta_3'(u)$ are defined by Eqs.~(\ref{def_delta}) and (\ref{def_delta3}), respectively.
The RG equation for $\delta(u)$ is given by
\begin{eqnarray}
\partial_l \delta(u) &=& - (D-3) \delta(u) \nonumber \\
&&+ l_2^+(z_l) \delta''(u) (\delta(0)-\delta(u))  \nonumber \\
&& -  l_2^-(z_l) \delta'(u)^2- 2 l_1^-(z_l) \delta_3'(u),
\label{RGeq_delta}
\end{eqnarray}
where $z_l$ is given by Eq.~(\ref{def_z}).
The RG equation for $\delta_3'(u)$ is presented in Appendix \ref{Appendix:Flow_equation_for_Delta_3}.

As in the equilibrium case, the critical exponent is related to the fixed point value of the dimensionless cumulant.
At a critical point, the correlation function for the transverse direction ($\bvec{r} \perp \bvec{e}_x$) decays as $C(\bvec{r}) = \overline{ \langle e^{i(u(r)-u(0))} \rangle } \sim |\bvec{r}|^{-\eta_{\perp}}$ with
\begin{equation}
\eta_{\perp} = \delta_*(0),
\label{eta_delta_neq}
\end{equation}
where $\delta_*(0)$ is the fixed point value of $\delta(0)$.
The derivation of Eq.~(\ref{eta_delta_neq}) is given in Appendix \ref{Appendix:Relation_between_exponent and_disorder}.

The flow equation of the $p$-th cumulant $\delta_p$ contains the $(p+1)$-th cumulant $\delta_{p+1}$.
Thus, we have an infinite hierarchy of the coupled flow equations.
However, in Eq.~(\ref{RGeq_delta}), the contribution of $\delta_3$ vanishes in the large length scale because $l_1^-(z_l) \sim e^{-2l}$.
In general, it can be shown that $\delta_{p+1}$ in the flow equation of $\delta_p$ always appears in the form of $l_1^-(z_l) \delta_{p+1}'$.
More precisely, $\Gamma_{p+1}$ appears in the exact flow equation for $\Gamma_p$ as
\begin{eqnarray}
&&\partial_l \Gamma_p[U_1,...,U_p] = \frac{n}{2} \mathrm{tr} \partial_l \mathbf{R}_k \mathrm{P}[U_1] \nonumber \\
&&\times \Gamma_{p+1}^{(110...0)}[U_1,U_1,U_2,...,U_p] \mathrm{P}[U_1],
\end{eqnarray}
and this term leads to $l_1^-(z_l) \delta_{p+1}'$.
This implies that the infinite hierarchy of the flow equations is decoupled in the large scale limit, which is rather surprising.
This conclusion does not rely on the specific functional form of $r (y_{\perp})$ in Eq.~(\ref{cutoff_function_neq}).
Note that such a decoupling does not occur in the equilibrium case (See Eq.~(\ref{RGeq_delta_eq})).

\subsection{RG evolution of the disorder cumulant}

We investigate the RG evolution of the disorder cumulant $\delta(u)$ at weak disorder.
The dimensional reduction predicts that the 4D-RFXYM and 3D-DRFXYM both exhibit QLRO with a continuously varying exponent in the weak disorder regime.
In the following, we will show that this prediction is true for the DRFXYM, but it is not for the RFXYM.

\subsubsection{Equilibrium case}
\label{sec:RG_evolution_Equilibrium_case}

For the equilibrium case ($v=0$), the FRG equation at $D=4$ is given by
\begin{eqnarray}
\partial_l \delta(u) = \delta''(u) (\delta(0)-\delta(u)) - \delta'(u)^2,
\label{RGeq_delta_eq_one_loop}
\end{eqnarray}
at weak disorder (See Eq.~(\ref{RGeq_delta_eq})).
If we assume that $\delta(u)$ is analytic at $u=0$, we have the flow equations for $\delta(0)$ and $\delta''(0)$ from Eq.~(\ref{RGeq_delta_eq_one_loop}),
\begin{eqnarray}
\partial_l \delta(0) &=& 0, \nonumber \\
\partial_l \delta''(0) &=& - 3\delta''(0)^2.
\end{eqnarray}
The first equation suggests the presence of a fixed line corresponding to QLRO.
However, from the second equation, $\delta''(0)$ diverges at a finite renormalization scale $l_L$, which is known as the Larkin scale \cite{Fisher-86}, and a linear cusp is generated at the origin, $\delta(u) \simeq \delta(0) + \delta'(0^+)|u|$.
This cusp leads to the non-trivial renormalization of $\delta(0)$,
\begin{eqnarray}
\partial_l \delta(0) = - \delta'(0^+)^2 \neq 0.
\label{flow_delta0_eq}
\end{eqnarray}
Such non-analytic behavior is a consequence of the presence of a large number of stationary states, as mentioned in the beginning of Sec.~\ref{Sec:RG_analysis_of_the_spin-wave_model}.
Note that, for an arbitrary $l$, $\delta_l(u)$ satisfies the potentiality condition,
\begin{equation}
\int^{2\pi}_0 \delta_l(u) du = 0.
\label{potentiality_condition}
\end{equation}

\begin{figure}
 \centering
 \includegraphics[width=0.45\textwidth]{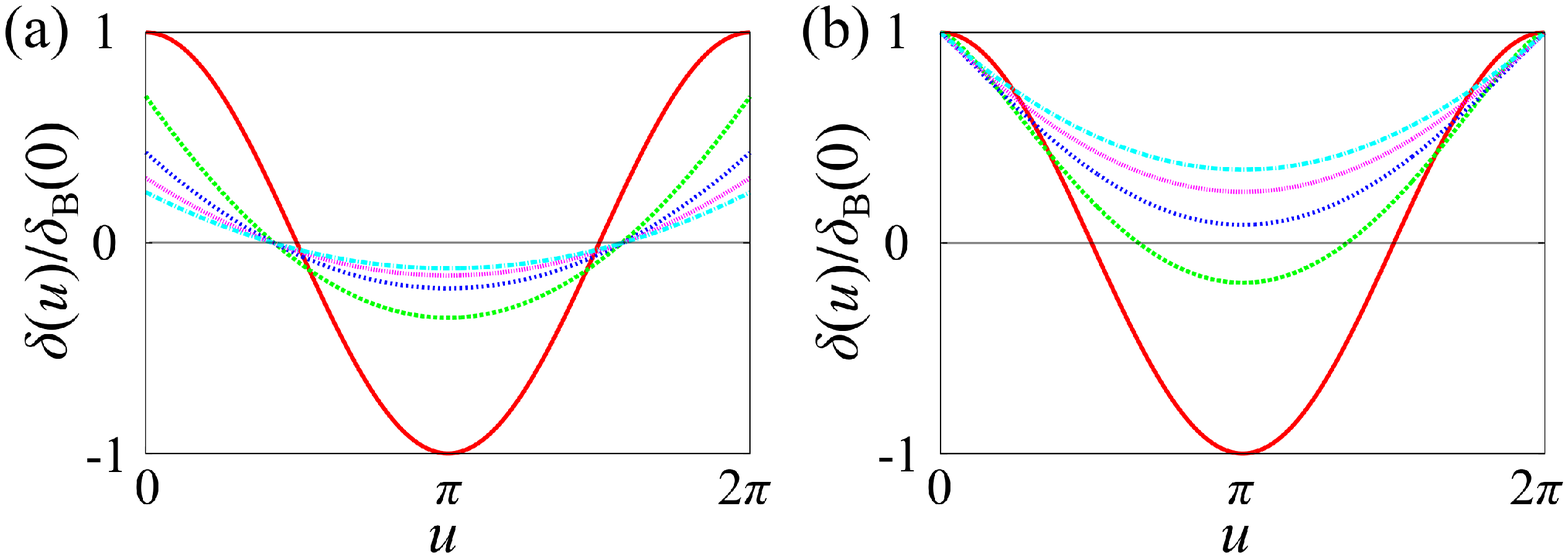}
 \caption{(a): RG evolution of $\delta(u)$ for the RFXYM at $D=4$ calculated by Eq.~(\ref{RGeq_delta_eq_one_loop}).
 The values of the renormalization scale are $l=0.0,1.0,2.0,3.0,4.0$ from the bottom to the top at $u=\pi$.
 The bare cumulant is given by the cosine function.
 $\delta(u)$ flows to zero in the limit $l \to \infty$.
 (b): RG evolution of $\delta(u)$ for the DRFXYM at $D=3$ calculated by Eq.~(\ref{RGeq_delta_large_scale}).
 $\delta(u)$ flows to a constant $\delta(u)=\delta_{\mathrm{B}}(0) \neq 0$ in the limit $l \to \infty$.}
 \label{fig:delta-evolution}
\end{figure}

The flow equation (\ref{RGeq_delta_eq_one_loop}) does not have any non-zero fixed point $\delta_*(u)$ satisfying Eq.~(\ref{potentiality_condition}).
This can be understood as follows.
From Eq.~(\ref{flow_delta0_eq}), $\delta_*'(0^+)=0$ at the fixed point, thus $\delta_*(u)$ does not have a linear cusp at the origin.
Next, note that $\delta_*''(u)$ should be non-negative for all $u$ because $\delta_*(0)-\delta_*(u)$ and $\delta_*'(u)^2$ are non-negative.
However, since $\delta_*(u)$ does not have a linear cusp, $\delta_*''(u)$ is negative near the origin.
Therefore, the fixed point satisfying Eq.~(\ref{potentiality_condition}) is only $\delta_*(u)=0$.
Since QLRO is characterized by a non-zero fixed point, the 4D-RFXYM does not exhibit QLRO with a continuously varying exponent.
Fig.~\ref{fig:delta-evolution} (a) shows the RG evolution of $\delta(u)$ obtained by numerical integration of Eq.~(\ref{RGeq_delta_eq_one_loop}).
The initial condition is given by the cosine function $\delta_{\mathrm{B}}(u)=h_0^2/(8\pi^2 K^2) \cos u$.
At the Larkin scale, it develops linear cusps at $u=2 \pi m$ and evolves into a parabolic profile.
Finally, $\delta(u)$ converges to zero because $\partial_l \delta(0)$ is negative (See Eq.~(\ref{flow_delta0_eq})).
Therefore, the disorder is marginally irrelevant.

We consider the phase structure of the 4D-RFXYM in detail.
We assume the following solution:
\begin{equation}
\delta_l(u)=\frac{\tilde{\delta}(u)}{l+l_0},
\end{equation}
where $\tilde{\delta}(u)$ is an $l$-independent function.
Substituting this expression into Eq.~(\ref{RGeq_delta_eq_one_loop}) yields
\begin{equation}
\tilde{\delta}(u) + \tilde{\delta}''(u)(\tilde{\delta}(0)-\tilde{\delta}(u)) - \tilde{\delta}'(u)^2 = 0,
\end{equation}
which can be solved as
\begin{equation}
\tilde{\delta}(u) = \frac{1}{6}(u-\pi)^2 - \frac{\pi^2}{18}, \:\:\: (0 \leq u < 2\pi).
\label{RFXYM_FP}
\end{equation}
Thus, we have $\delta_l(0) \simeq (\pi^2/9)l^{-1}$ for large $l$.
Since $\delta_l(0)$ corresponds to the scale-dependent exponent $\eta_l$ (See Eq.~(\ref{eta_delta_eq})), the correlation function satisfies
\begin{equation}
r \partial_r C(r) = - \delta_l(0) C(r) = - \frac{\pi^2}{9} (\ln r)^{-1} C(r),
\end{equation}
where we have used $l \sim \ln r$.
This leads to the following asymptotic behavior \cite{Feldman-00,Chitra-99}:
\begin{equation}
C(r) \sim (\ln r)^{-\pi^2/9}.
\end{equation}
Thus, the correlation function of the 4D-RFXYM decays more slowly than power-law.

\subsubsection{Nonequilibrium case}

We next consider the nonequilibrium case.
In the large length scale, $z_l$ in Eq.~(\ref{RGeq_delta}) can be set to zero. 
Thus, at $D=3$, we have the following FRG equation:
\begin{eqnarray}
\partial_l \delta(u) = \delta''(u) (\delta(0)-\delta(u)),
\label{RGeq_delta_large_scale}
\end{eqnarray}
which was first derived in Refs.~\cite{LeDoussal-98} and \cite{Balents-98} by using a perturbative approach at one-loop order.
Remarkably, from the decoupling nature mentioned in the end of Sec.~\ref{sec:flow_equations_neq}, Eq.~(\ref{RGeq_delta_large_scale}) is shown to be exact, at least within the ansatz Eqs.~(\ref{Gamma_1}) and (\ref{Gamma_n}).
The difference between Eqs.~(\ref{RGeq_delta_large_scale}) and (\ref{RGeq_delta_eq_one_loop}) is the absence of the last term $-\delta'(u)^2$.
If we assume that $\delta(u)$ is analytic at $u=0$, we have 
\begin{equation}
\partial_l \delta''(0) = - \delta''(0)^2.
\end{equation}
Therefore, as in the equilibrium case, $\delta''(0)$ diverges at a finite renormalization scale and a linear cusp is generated at the origin, $\delta(u) \simeq \delta(0) + \delta'(0^+)|u|$.

Fig.~\ref{fig:delta-evolution} (b) shows the RG evolution of $\delta(u)$ obtained by numerical integration of Eq.~(\ref{RGeq_delta_large_scale}).
The initial condition is given by the cosine function $\delta_{\mathrm{B}}(u)=h_0^2/(4\pi Kv) \cos u$.
At the Larkin scale, it develops linear cusps at $u=2 \pi m$ and evolves into a parabolic profile.
In contrast to the equilibrium case Eq.~(\ref{RGeq_delta_eq_one_loop}), it is important to note that $\delta(0)$ is not renormalized due to the absence of the term $-\delta'(u)^2$.
In the large scale limit $l \to \infty$, $\delta(u)$ eventually becomes a constant function $\delta_{l=\infty}(u) = \delta_{\mathrm{B}}(0)$.
Since the flow equations for the higher-order cumulants $\delta_p \: (p \geq 3)$ contain only the derivative of the same and lower-order cumulants $\delta_2,...,\delta_p$ in this limit, we can conclude that all higher-order cumulants vanish.
Therefore, there is a family of stable fixed points $\delta_{l=\infty}(u)=\mathrm{const.}$ and $\delta_3=\delta_4=...=0$.
These fixed points correspond to the massless free field theory in the sense that the renormalized random force is Gaussian and independent of the field $u$.
Then, the correlation function for the transverse direction ($\bvec{r} \perp \bvec{e}_x$) shows power-law decay,
$C(\bvec{r}) \sim |\bvec{r}|^{-\eta_{\perp}}$ with $\eta_{\perp}=\delta_{l=\infty}(0)$.
It is worth to note that the fixed points $\delta_{l=\infty}(u)=\mathrm{const.} \neq 0$ does not satisfy the potentiality condition Eq.~(\ref{potentiality_condition}) because a non-potential random force is generated due to the breakdown of the detailed balance condition.
The fact that $\delta_{p+1}$ in the flow equation of $\delta_p$ exponentially vanishes ensures that there is no other fixed point except the trivial one $\delta(u)=\mathrm{const.}$ and $\delta_3=\delta_4=...=0$.
In Eq.~(\ref{RGeq_delta_large_scale}), the terms $-l_2^-(z_l) \delta'(u)^2$ and $-2l_1^-(z_l) \delta_3'(u)$ in Eq.~(\ref{RGeq_delta}) are omitted from the beginning of its analysis.
Although these terms can slightly renormalize $\delta(0)$, they do not affect the stability of the fixed line discussed above.

Let us briefly remark on the differences between our study and the previous works such as Refs.~\cite{LeDoussal-98} and \cite{Balents-98}.
These authors derived the one-loop FRG equation (\ref{RGeq_delta_large_scale}) and discussed the transverse pinning of driven vortex lattices in dirty superconductors.
At zero temperature, the non-analytic behavior of the disorder cumulant $\delta(u)$ leads to trapping of the phase $u$, which corresponds to the transverse displacement in the context of the driven vortex lattices.
The critical force to ``depin'' the phase is proportional to the amplitude of the discontinuity in $\delta'(u)$ at the origin $u=0$.
The authors of Refs.~\cite{LeDoussal-98} and \cite{Balents-98} used Eq.~(\ref{RGeq_delta_large_scale}) to estimate this depinning critical force of the driven vortex lattices.
However, they did not mention the fact that this FRG equation has a line of fixed point at the critical dimension $D=3$, in contrast to the RFXYM the FRG equation of which has only a single trivial fixed point at $D=4$.
Actually, it is the presence of the fixed line in Eq.~(\ref{RGeq_delta_large_scale}) that our argument to characterize the topologically ordered phase is based on.
Furthermore, the non-perturbative FRG formalism enables us to estimate the higher-order contributions to Eq.~(\ref{RGeq_delta_large_scale}) and to show that they do not affect the stability of the fixed line.
Such an analysis is hard to perform within the perturbative approach presented in the previous works.

At weak disorder, the spin-wave model of the DRFXYM is found to show the QLRO in three dimensions.
For sufficiently strong disorder, we expect that the 3D-DRFXYM exhibits a transition to a disordered phase, wherein the correlation function $C(\bvec{r})$ decays exponentially.
However, it may be recalled that the spin-wave model is invalid in the strong disorder regime because it cannot describe the vortices.
The QLRO phase with a large $\eta_{\perp}$ in the spin-wave model is expected to be unstable against the vortices.
To describe the KT transition in the 3D-DRFXYM, we are required to consider the effect of vortices.

\section{Effects of the vortices}
\label{Sec:Effects_of_the_vortices}

We construct a phenomenological theory valid near the transition point by taking into account the effect of the vortices.
First, note that the dimensional reduction derived in Sec.~\ref{Sec:Dimensional_reduction} recovers in the large scale limit.
The criterion for the breakdown of the dimensional reduction is that the renormalized disorder cumulant corresponding to the fixed point has a linear cusp as a function of the field.
For the case of the spin-wave model, we have shown that $\delta_{l=\infty}(u)$ does not have any cusp because it is just a constant function.
This means that the dimensional reduction holds within the spin-wave approximation.
Since the amplitude fluctuations of the field, which are ignored in the spin-wave model, are irrelevant to the non-analytic behavior of the renormalized disorder cumulant, we expect that the dimensional reduction holds even when the vortices exist.
Therefore, the transverse section of the 3D-DRFXYM is identical to the 2D pure XY model.
This implies that the exponent $\eta_{\perp}$ should be equal to $1/4$ at the KT transition point.

This prediction is consistent with the result of our previous study Ref.~\cite{Haga-15}, wherein $\eta_{\perp}$ was determined by numerical simulation.
Fig.~1(c) in Ref.~\cite{Haga-15} shows $\eta_{\perp}$ as a function of the temperature for fixed disorder strength and driving velocity.
One can observe that at the transition temperature $\eta_{\perp}$ has a value close to $1/4$.
Although the argument based on the dimensional reduction is applicable only for the zero-temperature case, we expect that the value of the exponent $\eta_{\perp}$ at the transition point is universal even for finite-temperature case.

To construct a quantitative theory of the vortices for all disorder strength is too ambitious at this time.
Therefore, we attempt to establish the simplest theory that can be reduced to the spin-wave model in the weak disorder limit and that predicts $\eta_{\perp}=1/4$ at the transition point.
It provides a simple interpolation between the weak and strong disorder regimes.
To consider the effect of the vortices, we employ the following assumptions.
The first assumption is that the dominant effect of the vortices is to renormalize the elastic constant $K$.
This means that the phenomenological theory is obtained by replacing $K$ with $K_{\mathrm{eff}}$ in the spin-wave model, where $K_{\mathrm{eff}}$ is smaller than $K$.
For simplicity, we also assume that $K_{\mathrm{eff}}$ is isotropic.
The second assumption is that $K_{\mathrm{eff}}$ obeys the flow equation similar to that of the 2D pure XY model.
This assumption is a consequence of the recovery of the dimensional reduction.

We recall that the RG equations of the 2D pure XY model are given by
\begin{eqnarray}
\frac{d}{dl} \biggl( \frac{T}{K}\biggr) &=& 2 \pi^3 y^2,
\label{RG_2D_XY_K}
\end{eqnarray}
\begin{eqnarray}
\frac{dy}{dl} &=& \biggl( 2 - \pi \frac{K}{T} \biggr) y,
\label{RG_2D_XY_y}
\end{eqnarray}
where $y$ is the fugacity of the vortices, which is proportional to the vortex density \cite{Nelson-77}.
The dimensional reduction discussed in Sec.~\ref{Sec:Dimensional_reduction} implies that the disorder strength divided by the driving velocity in the 3D-DRFXYM corresponds to the temperature in the 2D pure XY model (See Eq.~(\ref{effective_temperature})).
Thus, the flow equation for $K_{\mathrm{eff}}$ can be obtained by replacing $T$ in Eq.~(\ref{RG_2D_XY_K}) with $\Delta(0)/(2v)$,
\begin{eqnarray}
\frac{\Delta(0)}{2v} \frac{d}{dl} \biggl( \frac{1}{K_{\mathrm{eff}}}\biggr) = 2 \pi^3 y^2.
\label{RGeq_Keff_intro}
\end{eqnarray}
In Eq.~(\ref{RGeq_Keff_intro}), the derivative operator $d/dl$ does not act on the dimensionful disorder $\Delta(0)$ because $K_{\mathrm{eff}}$ alone is modified by the vortices.
Note that $d K_{\mathrm{eff}}/dl = 0$ in the absence of the vortices $y=0$.
In terms of the dimensionless disorder $\delta(0)=\Delta(0)/(4 \pi K_{\mathrm{eff}}v)$, Eq.~(\ref{RGeq_Keff_intro}) can be rewritten as
\begin{eqnarray}
-\delta(0) \frac{d}{dl} \ln K_{\mathrm{eff}} = \pi^2 y^2.
\label{RGeq_Keff}
\end{eqnarray}
From Eq.~(\ref{RG_2D_XY_y}), the equation of the vortex fugacity can be obtained by the same replacement,
\begin{eqnarray}
\frac{dy}{dl} = \biggl( 2 - \frac{1}{2 \delta(0)} \biggr) y.
\label{RGeq_y}
\end{eqnarray}

Within the spin-wave approximation, the RG equations for $\delta(u)$ and $\delta_3'(u)$ are given by Eqs.~(\ref{RGeq_delta}) and (\ref{RGeq_delta3}) in Appendix \ref{Appendix:Flow_equation_for_Delta_3}, respectively.
To take into account the vortices, we replace $K$ in these flow equations with $K_{\mathrm{eff}}$,
\begin{eqnarray}
\partial_l \delta(u) &=& l_2^+(z_l) \delta''(u) (\delta(0)-\delta(u)) -  l_2^-(z_l) \delta'(u)^2 \nonumber \\
&&- 2 l_1^-(z_l) \delta_3'(u) -\delta(u) \frac{d}{dl} \ln K_{\mathrm{eff}},
\label{RGeq_delta_neq_vortex}
\end{eqnarray}
\begin{eqnarray}
\partial_l \delta_3'(u) &=& - 2 \delta_3'(u) - 2 l_3^+(z_l) \bigl[ \delta''(u) \delta'(u) (\delta(u)-\delta(0)) \bigr]' \nonumber \\
&&- l_3^-(z_l) \bigl[ \delta'(u)^3 - \delta'(0)^2 \delta'(u) \bigr]' \nonumber \\
&&- \delta_3'(u) \frac{d}{dl} \ln K_{\mathrm{eff}},
\label{RGeq_delta3_neq_vortex}
\end{eqnarray}
where $z_l=v^{-2} K_{\mathrm{eff}}^2 e^{-2l}$.
We have ignored the higher-order terms $\mathcal{O}(\delta_2 \delta_3)$ and $\mathcal{O}(\delta_4)$ in Eq.~(\ref{RGeq_delta3}) because they are expected to be small enough near the KT transition point.
The last terms proportional to $d \ln K_{\mathrm{eff}}/dl$ in the right-hand sides of Eqs.~(\ref{RGeq_delta_neq_vortex}) and (\ref{RGeq_delta3_neq_vortex}) come from the flow of $K_{\mathrm{eff}}$ in $\delta(u)=\Delta(u)/(4\pi K_{\mathrm{eff}} v)$ and $\delta_3'(u) = k^{2} \Delta_3'(u)/(16 \pi^2 K_{\mathrm{eff}} v^2)$.
The bare value of $y$ is also obtained from the replacement $T/K \to 2 \pi \delta_{\mathrm{B}}(0)$ in the 2D pure XY model \cite{Nelson-77}. 
Therefore, we have 
\begin{eqnarray}
y_0 = \exp \left[ -\frac{\pi}{4} \delta_{\mathrm{B}}(0)^{-1} \right].
\label{bare_y}
\end{eqnarray}
Eqs.~(\ref{RGeq_Keff}), (\ref{RGeq_y}), (\ref{RGeq_delta_neq_vortex}), and (\ref{RGeq_delta3_neq_vortex}) constitute a closed set of flow equations.

Unfortunately, Eq.~(\ref{RGeq_delta3_neq_vortex}) is numerically unstable due to the presence of the third derivative.
Since $\delta(u)$ and $\delta_{3}'(u)$ starting from any smooth bare cumulants eventually develop linear cusps in the large scale, it is convenient to consider a modified model whose bare cumulant $\delta_{\mathrm{B}}(u)$ already has a linear cusp at $u=0$.
We introduce the following functional form of the non-analytic cumulants:
\begin{eqnarray}
\delta_l(u) &=& a_1(l)(u-\pi)^2+b_1(l), \nonumber \\
\delta_{3,l}'(u) &=& a_2(l)(u-\pi)^2+b_2(l),
\label{RG_solution_form_neq}
\end{eqnarray}
for $u \in [0,2 \pi]$.
Eq.~(\ref{RG_solution_form_neq}) is extended to the whole region by using the periodicity, $\delta(u \pm 2\pi)=\delta(u)$ and $\delta_3'(u \pm 2\pi)=\delta_3'(u)$.
By substituting Eq.~(\ref{RG_solution_form_neq}) into Eqs.~(\ref{RGeq_delta_neq_vortex}) and (\ref{RGeq_delta3_neq_vortex}), we have
\begin{eqnarray}
\frac{d a_1}{dl} &=& -\bigl\{ 2 l_2^+(z_l) + 4  l_2^-(z_l) \bigr\} a_1^2 -  2 l_1^-(z_l) a_2 \nonumber \\
&&-a_1 \frac{d}{dl} \ln K_{\mathrm{eff}},
\label{RGeq_ab-1}
\end{eqnarray}
\begin{eqnarray}
\frac{d a_2}{dl} &=& -2 a_2 -24 \bigl\{ l_3^+(z_l) + l_3^-(z_l) \bigr\} a_1^3 \nonumber \\
&&-a_2 \frac{d}{dl} \ln K_{\mathrm{eff}},
\label{RGeq_ab-2}
\end{eqnarray}
\begin{eqnarray}
\frac{d}{dl} \delta(0) &=& - 4 \pi^2 l_2^-(z_l) a_1^2  - 2 l_1^-(z_l) \delta_3'(0) \nonumber \\
&&-\delta(0) \frac{d}{dl} \ln K_{\mathrm{eff}},
\label{RGeq_ab-3}
\end{eqnarray}
\begin{eqnarray}
\frac{d}{dl} \delta_3'(0) &=& - 2 \delta_3'(0) - 16 \pi^2 \bigl\{ l_3^+(z_l) + l_3^-(z_l) \bigr\} a_1^3 \nonumber \\
&&-\delta_3'(0) \frac{d}{dl} \ln K_{\mathrm{eff}}.
\label{RGeq_ab-4}
\end{eqnarray}
An anomalous term $\delta'(0)^2$ in Eq.~(\ref{RGeq_delta3_neq_vortex}), which vanishes for any analytic $\delta(u)$, has been calculated as $4 \pi^2 a_1^2$.
It is worth noting that the right-hand sides of Eqs.~(\ref{RGeq_delta_neq_vortex}) and (\ref{RGeq_delta3_neq_vortex}) neither yield cubic nor quartic terms of $u$ as the parabolic functions Eq.~(\ref{RG_solution_form_neq}) are substituted.
Thus, the set of equations (\ref{RGeq_ab-1})--(\ref{RGeq_ab-4}) is exact, provided Eqs.~(\ref{RGeq_delta_neq_vortex}), (\ref{RGeq_delta3_neq_vortex}), and (\ref{RG_solution_form_neq}).
We numerically solve this set of equations with the following initial condition: 
\begin{eqnarray}
\delta(0)&=&\delta_{\mathrm{B}}(0)=\frac{h_0^2}{4 \pi Kv}, \nonumber \\
\delta_3'(0)&=&0, \nonumber \\
a_1&=&\frac{3}{2\pi^2} \delta_{\mathrm{B}}(0), \nonumber \\
a_2&=&0,
\end{eqnarray}
where the initial value of $a_1$ is chosen such that the potentiality condition Eq.~(\ref{potentiality_condition}) is satisfied.

\begin{figure}
 \centering
 \includegraphics[width=0.45\textwidth]{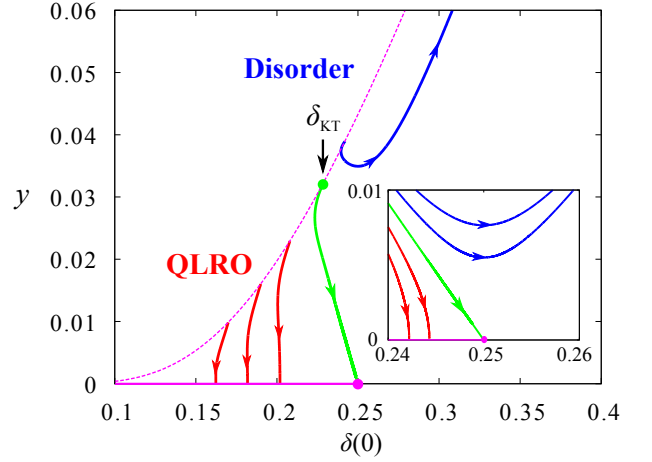}
 \caption{RG trajectories for the spin-wave model with the correction of the vortices.
 The horizontal axis is $\delta(0)$ and the vertical axis is $y$.
 The thin dashed curve is the bare value of the vortex fugacity Eq.~(\ref{bare_y}).
 The thick line on the horizontal axis is the fixed line characterizing the QLRO.
 The inset represents the trajectories near the endpoint of the fixed line.}
 \label{fig:RG-flow-KT}
\end{figure}

The RG trajectories with respect to the disorder and vortex fugacity are shown in Fig.~\ref{fig:RG-flow-KT}.
The trajectories starting from the weak disorder regime $\delta_{\mathrm{B}}(0) < \delta_{\mathrm{KT}}$ flow to the fixed line.
This regime corresponds to the QLRO phase.
In the weak disorder limit, the critical exponent $\eta_{\perp}=\delta_{l=\infty}(0)$ is equal to that predicted from the naive spin-wave calculation Eq.~(\ref{eta_SW}).
The trajectories starting from the strong disorder regime $\delta_{\mathrm{B}}(0) > \delta_{\mathrm{KT}}$ diverge $\delta(0), y \to \infty$.
This regime corresponds to a disordered phase.
At $\delta_{\mathrm{B}}(0)=\delta_{\mathrm{KT}}$, $\eta_{\perp} = 1/4$ as in the conventional KT transition.
Concerning $\eta_{\perp}=\delta_{l=\infty}(0)$ as a function of the bare disorder $\delta_{\mathrm{B}}(0)$, there are two distinct effects that lead to the deviation from Eq.~(\ref{eta_SW}).
The first one is the non-trivial renormalization of the disorder strength $\delta(0)$ resulting from the generation of the cusp in the renormalized cumulant.
This effect slightly lowers the value of $\eta_{\perp}$ than Eq.~(\ref{eta_SW}).
The second effect is the renormalization of the elastic constant due to the vortices.
Since the proliferation of the vortices leads to the decrease of the effective elastic constant, $\eta_{\perp}$ increases near the transition point $\delta_{\mathrm{KT}}$.
Although this theory may be unsatisfactory for the quantitative prediction of $\eta_{\perp}$, it is useful to understand the qualitative features of $\eta_{\perp}$ as a function of the bare disorder strength.

\begin{figure}
 \centering
 \includegraphics[width=0.4\textwidth]{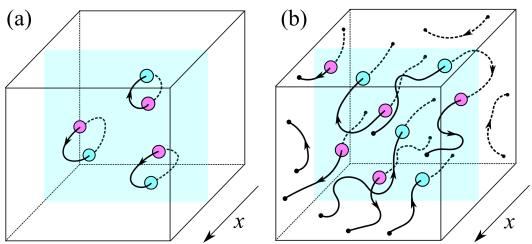}
 \caption{Schematic picture of the vortex dissociation process for the 3D-DRFXYM.
  The lines represent vortex lines.
  The transverse section corresponds to a snapshot of the 2D pure XY model.
  (a): QLRO phase. 
  The vortex lines form into small loops, whose volume density is low enough at weak disorder. 
  (b): Disordered phase.
  The space is filled with tangled long vortex lines.}
 \label{fig:vortex}
\end{figure}

Let us consider the changes in the vortex structure at the transition point.
The KT transition in the 2D XY model is understood as the dissociation of tightly bound vortex-antivortex pairs.
For the 3D-DRFXYM, this vortex dissociation picture should be modified because in three dimensions the vortex is a line, not a point.
According to the dimensional reduction property, in the large length scale, a snapshot of the 3D-DRFXYM is identical to a space-time trajectory of the 2D pure XY model by considering the spatial coordinate $x$ as a fictitious time.
In the QLRO phase of the 2D XY model, vortex-antivortex pairs are created by the thermal fluctuations, and they immediately annihilate by pair collisions.
Therefore, the QLRO phase of the 3D-DRFXYM is considered as a dilute gas of small vortex rings, which correspond to the space-time trajectories of the bound vortex-antivortex pairs.
The density of the vortex rings increases with the disorder strength, and eventually, at the transition point, such vortex rings merge into tangled long vortex lines, whose length is comparable to the system size.
The schematic picture is given in Fig.~\ref{fig:vortex}.

\section{Concluding remarks}
\label{Sec:Conclusion}

We have shown that the 3D-RFXYM exhibits QLRO and the KT transition when it is driven at a uniform and steady velocity.
In the first part of this paper, we have performed the FRG analysis of the spin-wave model of the DRFXYM and found that it flows to the massless free field theory in the large-scale limit.  
This fact ensures the existence of a line of fixed points characterizing QLRO with a continuously varying exponent.
Next, we have discussed the effect of the vortices, which leads to the breakdown of the QLRO in the strong disorder regime.
The main ingredient in this part is the dimensional reduction, which predicts that the transverse section of the 3D-DRFXYM is identical to the 2D XY model.
By using this property, we derive the flow equation for the effective elastic constant (helicity modulus) and the exponent $\eta_{\perp}$ is calculated as a function of the bare disorder strength.
We have also discussed the changes in the vortex structure at the KT transition.

In this study, the dimensional reduction plays a crucial role.
As mentioned in Sec.~\ref{Sec:Dimensional_reduction}, this property does not always hold.
In Ref.~\cite{Haga-17}, we investigated the critical behavior of the driven random field $O(N)$ model, which is defined by Eqs.~(\ref{Hamiltonian}) and (\ref{EM}) with an $N$-component vector field $\bvec{\phi}=(\phi^1,...,\phi^N)$.
The critical exponent $\eta$ was calculated as a function of $N$ in the first order of $\epsilon=D-3$ and we found that the dimensional reduction breaks down when $2<N<10$.
Therefore, the XY model ($N=2$) is an exceptional case where the dimensional reduction recovers.

This study is based on some assumptions and approximations the validity of which is not established yet.
In the first part of this paper wherein the FRG analysis of the spin-wave model was performed, we have employed the simplest approximations for the functional forms of the effective action, Eqs.~(\ref{Gamma_1}), (\ref{Gamma_n}) and that of the cutoff function, Eq.~(\ref{cutoff_function_neq}).
These approximations may be justified by the fact that, in the leading order of the disorder, the resulting flow equation (\ref{RGeq_delta_large_scale}) is the same as that obtained from the perturbative approach in the previous studies \cite{LeDoussal-98,Balents-98}.
In the second part wherein the effect of the vortices was discussed, the critical assumption is that the dimensional reduction derived in Sec.~\ref{Sec:Dimensional_reduction} holds if the renormalized cumulant is analytic as a function of the field at the fixed point.
The relation between the non-analytic nature of the disorder cumulant and the breakdown of the dimensional reduction is well established for the equilibrium cases, but it is not yet for the nonequilibrium cases discussed in this paper.
As an evidence supporting this assumption, we refer to our previous work Ref.~\cite{Haga-17}, where we confirmed that in the driven random field $O(N)$ model the dimensional reduction fails if and only if the renormalized disorder cumulant has a cusp.
Unfortunately, at this time there is no general argument to justify this assumption.
Therefore, one should keep in mind that our results are based on non-controlled approximations and further theoretical and numerical investigations are required.

One of the strategies to numerically verify the 3D KT transition is to investigate the behavior of the helicity modulus $K_{\mathrm{eff}}$.
In numerical simulations, it can be determined by measuring the force required to slightly twist the spins at the boundary of the 3D simulation box.
For second order transitions, the helicity modulus vanishes continuously at the critical point, while for the KT transition it exhibits a discontinuous jump.
Thus, one can distinguish between second order transition and KT transition from the behavior of the helicity modulus.
Furthermore, in the conventional KT transition in the 2D XY model, the amplitude of the jump in the helicity modulus satisfies a universal relation \cite{Nelson-77},
\begin{equation}
\frac{K_{\mathrm{eff}}}{T} = \frac{2}{\pi}.
\label{univ_jump_2D}
\end{equation}
We can also derive such a relation for the 3D KT transition.
If the renormalization of the disorder $\Delta(0)$ is negligible, the universal relation for the helicity modulus is obtained by replacing the temperature $T$ in Eq.~(\ref{univ_jump_2D}) with $h_0^2/(2v)$,
\begin{equation}
\frac{K_{\mathrm{eff}} v}{h_0^2} = \frac{1}{\pi}.
\label{univ_jump}
\end{equation}
This relation can be also obtained from the fact that the exponent $\eta_{\perp}=h_0^2/(4\pi K_{\mathrm{eff}} v)$ is equal to $1/4$ at the transition point.
Eq.~(\ref{univ_jump}) can be easily checked by simulations.
It is also interesting to numerically investigate the scenario depicted in Fig.~\ref{fig:vortex}.
As the disorder strength increases through the transition point, one may observe that the small vortex rings turn into tangled long vortex lines.
These numerical studies will be reported in future publications.

\begin{acknowledgments}
We are grateful to Gilles Tarjus and Matthieu Tissier for their many enlightening discussions and remarks.
We also acknowledge Shin-ich Sasa for our useful discussions.
The present study was supported by JSPS KAKENHI Grant No. 15J01614, a Grant-in-Aid for JSPS Fellows.
\end{acknowledgments}

\appendix

\section{Derivation of the dynamical action}
\label{Appendix:Derivation_of_the_dynamical_action}

In this Appendix, we derive the dynamical action of the spin-wave model, Eq.~(\ref{Bare_action}).
We introduce an $n$-replicated system with the same disorder,
\begin{eqnarray}
\partial_t u_a + v \partial_x u_a = K \nabla^2 u_a + F(\bvec{r};u_a) + \xi_a(\bvec{r},t),
\label{EM_SW_replicated}
\end{eqnarray}
where the subscript $a=1,...,n$ is the replica index.
The thermal noise satisfies
\begin{equation}
\langle \xi_a(\bvec{r},t) \xi_b(\bvec{r}',t') \rangle =2 T \delta_{ab} \delta(\bvec{r}-\bvec{r'}) \delta(t-t').
\end{equation}
For an arbitrary functional $A[\{u_a\}]$, its average over the thermal noise is written as
\begin{eqnarray}
\langle A[\{u_a\}] \rangle = \int \mathcal{D} \xi P[\xi] \int \mathcal{D} u A[\{u_a\}] \delta(u_a-u_a[\xi]) ,
\end{eqnarray}
where $u_a[\xi]$ is the solution of Eq.~(\ref{EM_SW_replicated}) for a realization of the noise $\xi_a$ and $P[\xi]$ is the probability distribution function of $\xi_a$.
This average can be calculated as
\begin{eqnarray}
\langle A[\{u_a\}] \rangle &=& \int \mathcal{D} \xi P[\xi] \int \mathcal{D} u A[\{u_a\}] \delta \bigl[ \partial_t u_a + v \partial_x u_a \nonumber \\
&& - K \nabla^2 u_a - F(\bvec{r};u_a) - \xi_a \bigr] \nonumber \\
&=& \int \mathcal{D} \xi P[\xi] \int \mathcal{D} u \mathcal{D} \hat{u} A[\{u_a\}] \exp \biggl[ - \sum_a \nonumber \\
&& \int_{rt} i \hat{u}_a \bigl\{ \partial_t u_a + v \partial_x u_a - K \nabla^2 u_a \nonumber \\
&&- F(\bvec{r};u_a) - \xi_a \bigr\} \biggr] \nonumber \\
&=& \int \mathcal{D} u \mathcal{D} \hat{u} A[\{u_a\}] \exp \biggl[ - \sum_a \nonumber \\
&& \int_{rt} i \hat{u}_a \bigl\{ \partial_t u_a + v \partial_x u_a - K \nabla^2 u_a\nonumber \\
&& - F(\bvec{r};u_a) - T i \hat{u}_a \bigr\} \biggr],
\end{eqnarray}
where the Jacobian associated with the delta function can be set to unity \cite{Canet-11}.
The disorder average is given by
\begin{eqnarray}
\overline{\langle A[\{u_a\}] \rangle} = \int \mathcal{D} U A[\{u_a\}] \exp \bigl( -S[\{ U_a \}] \bigr),
\end{eqnarray}
\begin{eqnarray}
S[\{ U_a \}] &=& \sum_a \int_{rt} i\hat{u}_a \bigl[ \partial_t u_a - T i\hat{u}_a + v \partial_x u_a - K \nabla^2 u_a \bigr] \nonumber \\
&&- \frac{1}{2} \sum_{a,b} \int_{rtt'} i\hat{u}_{a,t} i\hat{u}_{b,t'} \Delta_{\mathrm{B}}( u_{a,t} - u_{b,t'} ).
\end{eqnarray}
To simplify the notation, we have omitted the imaginary unit $i$ in the main text.

\section{Exact flow equations for $\Gamma_{p,k}$}
\label{Appendix:Exact_flow_equations}

In this Appendix, we derive the exact flow equations for $\Gamma_{p,k}$.
We omit the subscript $k$ in $\Gamma_{p,k}$ below.
The exact flow equation for $\Gamma$ is given by Eq.~(\ref{exact_flow_equation}).
To calculate the inverse of $\Gamma^{(2)} + \hat{\mathbf{R}}_k$ with respect to the replica indices, we rewrite it as
\begin{eqnarray}
\bigl( \Gamma^{(2)} + \hat{\mathbf{R}}_k \bigr)_{ab} &=& \mathrm{P}[U_a]^{-1} \delta_{ab} - A[U_a] \delta_{ab} \nonumber \\
&&- B[U_a,U_b],
\label{Appendix_Gamma(2)_R}
\end{eqnarray}
where $\mathrm{P}[U]$ is the one-replica propagator defined by
\begin{equation}
\mathrm{P}[U] = \bigl[ \Gamma_{1}^{(2)}[U] + \mathbf{R}_k \bigr]^{-1}.
\end{equation}
$A[U_a]$ and $B[U_a,U_b]$ can be expanded by increasing the number of free replica sums,
\begin{eqnarray}
A[U_a] &=& \sum_c A^{[1]}[U_a|U_c] \nonumber \\
&&+ \frac{1}{2} \sum_{c,d} A^{[2]}[U_a|U_c,U_d] +...,
\label{Appendix_A_expansion}
\end{eqnarray}
\begin{eqnarray}
B[U_a,U_b] &=& B^{[0]}[U_a,U_b] + \sum_c B^{[1]}[U_a,U_b|U_c] \nonumber \\
&&+ \frac{1}{2} \sum_{c,d} B^{[2]}[U_a,U_b|U_c,U_d] + ...,
\label{Appendix_B_expansion}
\end{eqnarray}
where the vertical bar in each term $A^{[p]}[U_a|U_{c_1},...,U_{c_p}]$ is introduced to distinguish between the ``explicit'' index $a$ and the dummy indices $c_1,...,c_p$, which run from $1$ to $n$ as the summation is taken.
In the following, we use simplified notations such as,
\begin{eqnarray}
\Gamma_3^{(200)}[U_1,U_2,U_3] = \frac{\delta^2 \Gamma_3[U_1,U_2,U_3]}{\delta U_1 \delta U_1}, \nonumber \\
\Gamma_3^{(110)}[U_1,U_2,U_3] = \frac{\delta^2 \Gamma_3[U_1,U_2,U_3]}{\delta U_1 \delta U_2}.
\end{eqnarray}
From Eq.~(\ref{Gamma_expansion}), $A^{[p]}$ and $B^{[p]}$ are written as
\begin{eqnarray}
A^{[1]}[U_a|U_c] &=& \Gamma_2^{(20)}[U_a,U_c], \nonumber \\
A^{[2]}[U_a|U_c,U_d] &=& - \Gamma_3^{(200)}[U_a,U_c,U_d],
\label{Appendix_A_Gamma}
\end{eqnarray}
and
\begin{eqnarray}
B^{[0]}[U_a,U_b] &=& \Gamma_2^{(11)}[U_a,U_b], \nonumber \\
B^{[1]}[U_a,U_b|U_c] &=& - \Gamma_3^{(110)}[U_a,U_b,U_c], \nonumber \\
B^{[2]}[U_a,U_b|U_c,U_d] &=& \Gamma_4^{(1100)}[U_a,U_b,U_c,U_d].
\label{Appendix_B_Gamma}
\end{eqnarray}
The inverse of Eq.~(\ref{Appendix_Gamma(2)_R}) reads
\begin{eqnarray}
&&\bigl( \Gamma^{(2)} + \hat{\mathbf{R}}_k \bigr)^{-1}_{ab} = \mathrm{P}[U_a] \delta_{ab} \nonumber \\
&&+ \mathrm{P}[U_a]\bigl( A[U_a] \delta_{ab}+B[U_a,U_b]\bigr)\mathrm{P}[U_b] \nonumber \\
&&+ \mathrm{P}[U_a]\bigl( A[U_a] \delta_{ac}+B[U_a,U_c] \bigr)\mathrm{P}[U_c] \nonumber \\
&&\times \bigl( A[U_c] \delta_{cb}+B[U_c,U_b] \bigr)\mathrm{P}[U_b] +...,
\label{propagator_expansion}
\end{eqnarray}
where the summation over repeated indices is assumed.
By substituting Eqs.~(\ref{Appendix_A_expansion}) and (\ref{Appendix_B_expansion}) into (\ref{propagator_expansion}), and taking the trace over the replica indices, we have the exact flow equations for $\Gamma_p$.

\begin{figure}
 \centering
 \includegraphics[width=0.25\textwidth]{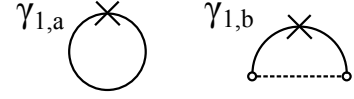}
 \caption{Graphical representation of the flow equation for $\Gamma_1$.}
 \label{fig:gamma-1}
\end{figure}

The exact flow equation for $\Gamma_1$ reads
\begin{eqnarray}
\partial_l \Gamma_1[U] = \frac{1}{2} \gamma_{1,a} + \frac{1}{2} \gamma_{1,b},
\label{Appendix_gamma_1}
\end{eqnarray}
\begin{eqnarray}
\gamma_{1,a} &=& \mathrm{tr} \partial_l \mathbf{R}_k \mathrm{P}[U], \nonumber
\end{eqnarray}
\begin{eqnarray}
\gamma_{1,b} &=& \mathrm{tr} \partial_l \mathbf{R}_k \mathrm{P}[U] \Gamma_2^{(11)}[U,U] \mathrm{P}[U], \nonumber
\end{eqnarray}
where ``$\mathrm{tr}$'' represents an integration over momentum and frequency as well as a sum over two conjugate fields $\{ u,\hat{u} \}$.
Fig.~\ref{fig:gamma-1} shows the graphical representation of the flow equation for $\Gamma_1$.
The rule for the graphical representation is as follows:
\begin{enumerate}
 \item An inner line denotes the propagator $\mathrm{P}[U]$.
 \item Two open dots linked by a dashed line represent a vertex obtained from a derivative of the two-replica action $\Gamma_2^{(p_1 p_2)}[U_1,U_2]$.
 \item A cross symbol denotes $\partial_l \mathbf{R}_k$.
\end{enumerate}

\begin{figure}
 \centering
 \includegraphics[width=0.4\textwidth]{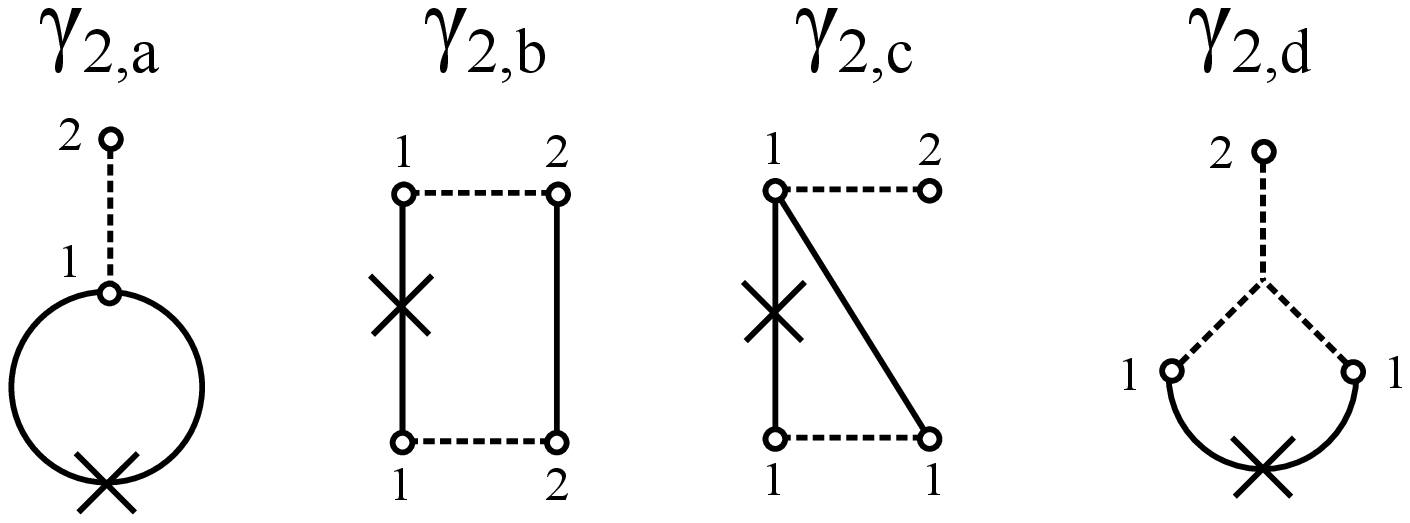}
 \caption{Graphical representation of the flow equation for $\Gamma_2$.}
 \label{fig:gamma-2}
\end{figure}

The exact flow equation for $\Gamma_2$ reads
\begin{eqnarray}
\partial_l \Gamma_2[U_1,U_2] &=& -\frac{1}{2} \bigl[ \gamma_{2,a} + \gamma_{2,b} + 2 \gamma_{2,c} - \gamma_{2,d} \nonumber \\
&&+ \mathrm{perm} \bigr],
\label{Appendix_gamma_2}
\end{eqnarray}
\begin{eqnarray}
\gamma_{2,a} &=& \mathrm{tr} \partial_l \mathbf{R}_k \mathrm{P}[U_1] \Gamma_2^{(20)}[U_1,U_2] \mathrm{P}[U_1], \nonumber
\end{eqnarray}
\begin{eqnarray}
\gamma_{2,b} &=& \mathrm{tr} \partial_l \mathbf{R}_k \mathrm{P}[U_1] \Gamma_2^{(11)}[U_1,U_2] \mathrm{P}[U_2] \nonumber \\
&&\times \Gamma_2^{(11)}[U_2,U_1] \mathrm{P}[U_1], \nonumber
\end{eqnarray}
\begin{eqnarray}
\gamma_{2,c} &=& \mathrm{tr} \partial_l \mathbf{R}_k \mathrm{P}[U_1] \Gamma_2^{(20)}[U_1,U_2] \mathrm{P}[U_1] \nonumber \\ 
&&\times \Gamma_2^{(11)}[U_1,U_1] \mathrm{P}[U_1], \nonumber
\end{eqnarray}
\begin{eqnarray}
\gamma_{2,d} &=& \mathrm{tr} \partial_l \mathbf{R}_k \mathrm{P}[U_1] \Gamma_3^{(110)}[U_1,U_1,U_2] \mathrm{P}[U_1]. \nonumber
\end{eqnarray}
Fig.~\ref{fig:gamma-2} shows the graphical representation of the flow equation for $\Gamma_2$.

\begin{figure}
 \centering
 \includegraphics[width=0.4\textwidth]{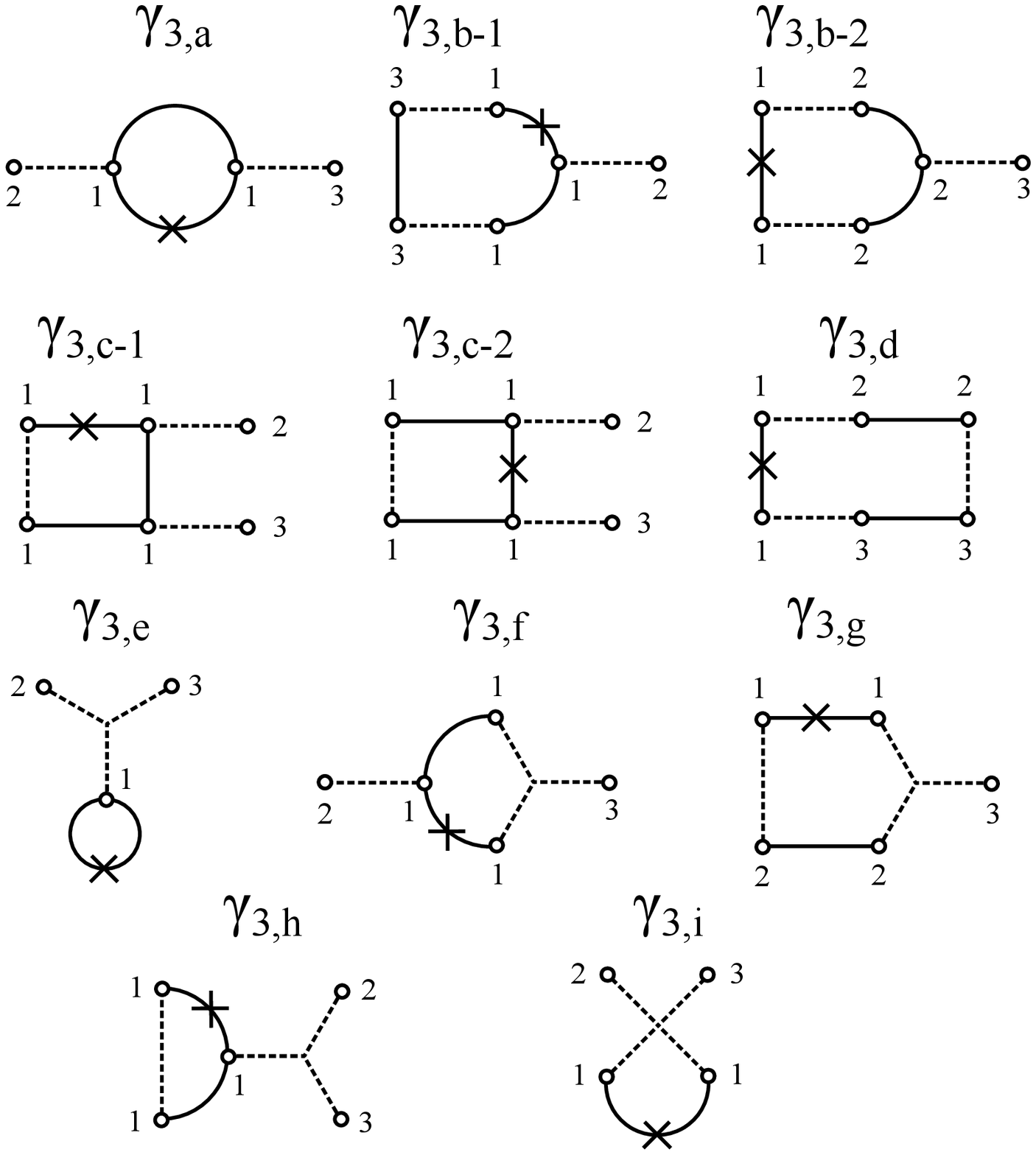}
 \caption{Graphical representation of the flow equation for $\Gamma_3$.}
 \label{fig:gamma-3}
\end{figure}

The exact flow equation for $\Gamma_3$ reads
\begin{eqnarray}
\partial_l \Gamma_3[U_1,U_2,U_3] &=& \frac{1}{2} \bigl[ \gamma_{3,a} + 2\gamma_{3,b-1} + \gamma_{3,b-2} \nonumber \\
&&+ 2\gamma_{3,c-1} + \gamma_{3,c-2} + \gamma_{3,d} \nonumber \\
&&- \gamma_{3,e} - 2\gamma_{3,f} - 2\gamma_{3,g} \nonumber \\
&&- \gamma_{3,h} + \gamma_{3,i} + \mathrm{perm} \bigr],
\label{Appendix_gamma_3}
\end{eqnarray}
\begin{eqnarray}
\gamma_{3,a} &=& \mathrm{tr} \partial_l \mathbf{R}_k \mathrm{P}[U_1] \Gamma_2^{(20)}[U_1,U_2] \mathrm{P}[U_1] \nonumber \\
&&\times \Gamma_2^{(20)}[U_1,U_3] \mathrm{P}[U_1], \nonumber
\end{eqnarray}
\begin{eqnarray}
\gamma_{3,b-1} &=& \mathrm{tr} \partial_l \mathbf{R}_k \mathrm{P}[U_1] \Gamma_2^{(20)}[U_1,U_2] \mathrm{P}[U_1] \Gamma_2^{(11)}[U_1,U_3] \nonumber \\
&&\times \mathrm{P}[U_3] \Gamma_2^{(11)}[U_3,U_1] \mathrm{P}[U_1], \nonumber
\end{eqnarray}
\begin{eqnarray}
\gamma_{3,b-2} &=& \mathrm{tr} \partial_l \mathbf{R}_k \mathrm{P}[U_1] \Gamma_2^{(11)}[U_1,U_2] \mathrm{P}[U_2] \Gamma_2^{(20)}[U_2,U_3] \nonumber \\
&&\times \mathrm{P}[U_2] \Gamma_2^{(11)}[U_2,U_1] \mathrm{P}[U_1], \nonumber
\end{eqnarray}
\begin{eqnarray}
\gamma_{3,c-1} &=& \mathrm{tr} \partial_l \mathbf{R}_k \mathrm{P}[U_1] \Gamma_2^{(20)}[U_1,U_2] \mathrm{P}[U_1] \Gamma_2^{(20)}[U_1,U_3] \nonumber \\
&&\times \mathrm{P}[U_1] \Gamma_2^{(11)}[U_1,U_1] \mathrm{P}[U_1], \nonumber
\end{eqnarray}
\begin{eqnarray}
\gamma_{3,c-2} &=& \mathrm{tr} \partial_l \mathbf{R}_k \mathrm{P}[U_1] \Gamma_2^{(20)}[U_1,U_2] \mathrm{P}[U_1] \Gamma_2^{(11)}[U_1,U_1] \nonumber \\
&&\times \mathrm{P}[U_1] \Gamma_2^{(20)}[U_1,U_3] \mathrm{P}[U_1], \nonumber 
\end{eqnarray}
\begin{eqnarray}
\gamma_{3,d} &=& \mathrm{tr} \partial_l \mathbf{R}_k \mathrm{P}[U_1] \Gamma_2^{(11)}[U_1,U_2] \mathrm{P}[U_2] \Gamma_2^{(11)}[U_2,U_3] \nonumber \\
&&\times \mathrm{P}[U_3] \Gamma_2^{(11)}[U_3,U_1] \mathrm{P}[U_1], \nonumber
\end{eqnarray}
\begin{eqnarray}
\gamma_{3,e} &=& \frac{1}{2} \mathrm{tr} \partial_l \mathbf{R}_k \mathrm{P}[U_1] \Gamma_3^{(200)}[U_1,U_2,U_3] \mathrm{P}[U_1], \nonumber 
\end{eqnarray}
\begin{eqnarray}
\gamma_{3,f} &=& \mathrm{tr} \partial_l \mathbf{R}_k \mathrm{P}[U_1] \Gamma_2^{(20)}[U_1,U_2] \mathrm{P}[U_1] \nonumber \\
&&\times \Gamma_3^{(110)}[U_1,U_1,U_3] \mathrm{P}[U_1], \nonumber
\end{eqnarray}
\begin{eqnarray}
\gamma_{3,g} &=& \mathrm{tr} \partial_l \mathbf{R}_k \mathrm{P}[U_1] \Gamma_2^{(11)}[U_1,U_2] \mathrm{P}[U_2] \nonumber \\ 
&&\times \Gamma_3^{(110)}[U_2,U_1,U_3] \mathrm{P}[U_1], \nonumber
\end{eqnarray}
\begin{eqnarray}
\gamma_{3,h} &=& \mathrm{tr} \partial_l \mathbf{R}_k \mathrm{P}[U_1] \Gamma_2^{(11)}[U_1,U_1] \mathrm{P}[U_1] \nonumber \\
&&\times \Gamma_3^{(200)}[U_1,U_2,U_3] \mathrm{P}[U_1], \nonumber
\end{eqnarray}
\begin{eqnarray}
\gamma_{3,i} &=& \frac{1}{2} \mathrm{tr} \partial_l \mathbf{R}_k \mathrm{P}[U_1] \Gamma_4^{(1100)}[U_1,U_1,U_2,U_3] \mathrm{P}[U_1]. \nonumber
\end{eqnarray}
Fig.~\ref{fig:gamma-3} shows the graphical representation of the flow equation for $\Gamma_3$.

\section{Flow equation for $\Delta_3$}
\label{Appendix:Flow_equation_for_Delta_3}

In this Appendix, we present the flow equation for the third cumulant $\Delta_3$.
From Eqs.~(\ref{Appendix_gamma_3}), (\ref{Gamma_1}), (\ref{Gamma_n}), and (\ref{RGeq_Delta-0}), the flow equation is given as follows:
\begin{eqnarray}
\partial_l \Delta_3(u_1,u_2,u_3) &=& \mathrm{A}_1 + ... + \mathrm{A}_5 \nonumber \\
&&+ \mathrm{B}_1 + ... + \mathrm{B}_6 \nonumber \\
&&+ \mathrm{C}_1 + \mathrm{perm},
\label{RG_Delta3_dimensionfull}
\end{eqnarray}
\begin{eqnarray}
\mathrm{A}_1 = -\Delta_2^{(20)}(u_1,u_2) \Delta_2^{(01)}(u_1,u_3) \Delta_2(u_1,u_3) L_3^+, \nonumber
\end{eqnarray}
\begin{eqnarray}
\mathrm{A}_2 = -\Delta_2^{(10)}(u_1,u_2) \Delta_2^{(11)}(u_1,u_3) \Delta_2(u_1,u_3) L_3^+, \nonumber
\end{eqnarray}
\begin{eqnarray}
\mathrm{A}_3 = -\Delta_2^{(10)}(u_1,u_2) \Delta_2^{(10)}(u_1,u_3) \Delta_2^{(01)}(u_1,u_3) L_3^-, \nonumber
\end{eqnarray}
\begin{eqnarray}
\mathrm{A}_4 = -\Delta_2^{(20)}(u_1,u_2) \Delta_2^{(10)}(u_1,u_3) \Delta_2(u_1,u_1) L_3^+, \nonumber
\end{eqnarray}
\begin{eqnarray}
\mathrm{A}_5 = -\Delta_2^{(11)}(u_1,u_2) \Delta_2^{(01)}(u_2,u_3) \Delta_2(u_1,u_3) L_3^+, \nonumber
\end{eqnarray}
\begin{eqnarray}
\mathrm{B}_1 = \frac{1}{2} \Delta_2^{(20)}(u_1,u_2) \Delta_3(u_1,u_1,u_3) L_2^+, \nonumber
\end{eqnarray}
\begin{eqnarray}
\mathrm{B}_2 = \Delta_2^{(10)}(u_1,u_2) \Delta_3^{(100)}(u_1,u_1,u_3) L_2^-, \nonumber
\end{eqnarray}
\begin{eqnarray}
\mathrm{B}_3 = \frac{1}{2} \Delta_2^{(11)}(u_1,u_2) \Delta_3(u_1,u_2,u_3) L_2^+, \nonumber
\end{eqnarray}
\begin{eqnarray}
\mathrm{B}_4 = \Delta_2^{(10)}(u_1,u_2) \Delta_3^{(010)}(u_1,u_2,u_3) L_2^-, \nonumber
\end{eqnarray}
\begin{eqnarray}
\mathrm{B}_5 = \frac{1}{2} \Delta_2(u_1,u_2) \Delta_3^{(110)}(u_1,u_2,u_3) L_2^+, \nonumber
\end{eqnarray}
\begin{eqnarray}
\mathrm{B}_6 = \frac{1}{4} \Delta_2(u_1,u_1) \Delta_3^{(200)}(u_1,u_2,u_3) L_2^+, \nonumber
\end{eqnarray}
\begin{eqnarray}
\mathrm{C}_1 = - \frac{1}{2} \Delta_4^{(1000)} (u_1,u_1,u_2,u_3) L_1^-, \nonumber
\end{eqnarray}
where ``$\mathrm{perm}$'' denotes the expressions obtained by permuting $u_1$, $u_2$, and $u_3$, and $L_n^{\pm}$ is defined by Eq.~(\ref{def_L}).

\subsection{Equilibrium case}

For the equilibrium case ($v=0$), the dimensionless cumulants are defined by Eqs.~(\ref{def_delta_eq}) and (\ref{def_delta3_eq}).
If we define $\delta_3'(u)$ by Eq.~(\ref{def_delta3}), its flow equation is given by
\begin{eqnarray}
\partial_l \delta_3'(u) &=& - (2D-6) \delta_3'(u) \nonumber \\
&&- \frac{3}{2} \bigl[ (\delta(u)-\delta(0)) \delta'(u)^2- \delta'(0)^2 \delta(u) \bigr]''  \nonumber \\
&&+\mathcal{O}(\delta_2 \delta_3) +\mathcal{O}(\delta_4).
\label{RGeq_delta3_eq}
\end{eqnarray}
Note that an ``anomalous'' term $\delta'(0)^2$ in Eq.~(\ref{RGeq_delta3_eq}) vanishes if $\delta(u)$ is analytic at $u=0$.
Since $\delta_2 \delta_3=\mathcal{O}(\delta_2^4)$ and $\delta_4=\mathcal{O}(\delta_2^4)$, the last terms in Eq.~(\ref{RGeq_delta3_eq}) can be negligible at weak disorder.

If we consider a fixed point at $D=4+\epsilon$, $\delta_3'(u)$ in Eq.~(\ref{RGeq_delta_eq}) can be eliminated by using Eq.~(\ref{RGeq_delta3_eq}), and finally we have
\begin{eqnarray}
0 &=& - \epsilon \delta(u) + \delta''(u) (\delta(0)-\delta(u)) - \delta'(u)^2 \nonumber \\
&&+ C \bigl[ (\delta(u)-\delta(0)) \delta'(u)^2 - \delta'(0)^2 \delta(u) \bigr]'',
\label{RGeq_delta_2-loop_eq}
\end{eqnarray}
where the constant $C$ depends on the choice of the cutoff function $R_k(\mathbf{q})$.
In the case of the optimized cutoff function Eq.~(\ref{optimized_cutoff_function}), $C=3/4$.
Eq.~(\ref{RGeq_delta_2-loop_eq}) can be also obtained from the two-loop perturbative FRG calculation \cite{LeDoussal-04}, however, $C=1/2$ in this case.

\subsection{Nonequilibrium case}

For the nonequilibrium case, the dimensionless cumulants are defined by Eqs.~(\ref{def_delta_neq}) and (\ref{def_delta3_neq}).
The flow equation $\delta_3'(u)$ is then given by
\begin{eqnarray}
\partial_l \delta_3'(u) &=& - (2D-4) \delta_3'(u) \nonumber \\
&&- 2 l_3^+(z_l) \bigl[ \delta''(u) \delta'(u) (\delta(u)-\delta(0)) \bigr]' \nonumber \\
&&- l_3^-(z_l) \bigl[ \delta'(u)^3 - \delta'(0)^2 \delta'(u) \bigr]' \nonumber \\
&&+ \mathcal{O}(\delta_2 \delta_3) + \mathcal{O}(\delta_4),
\label{RGeq_delta3}
\end{eqnarray}
where $l_n^-(z)$ and $l_n^+(z)$ are defined by Eqs.~(\ref{def_l_n-}) and (\ref{def_l_n+}).

\section{Relation between the critical exponent and the renormalized disorder}
\label{Appendix:Relation_between_exponent and_disorder}

\subsection{Equilibrium case}

We show that, at a critical point, the exponent of the correlation function $\eta$ is equal to the dimensionless renormalized disorder $\delta_*(0)$ at the fixed point, Eq.~(\ref{eta_delta_eq}).
First, note that, at zero temperature, the correlation of the phase parameter $u$ is given by
\begin{equation}
\overline{\langle u(\mathbf{q}) u(-\mathbf{q}) \rangle} = \frac{\Delta(0)}{K^2 |\mathbf{q}|^4},
\end{equation}
where $\Delta(0)$ is the dimensionfull disorder strength.
From Eq.~(\ref{def_delta_eq}), $\Delta(0)$ scales as $\Delta(0)=8\pi^2 K^2 k^{\epsilon} \delta_*(0)$ at $D=4-\epsilon$, where the prefactor $(16/D)A_D$ is evaluated at $D=4$.
If the running scale $k$ is replaced with $|\mathbf{q}|$, we have
\begin{equation}
\overline{\langle u(\mathbf{q}) u(-\mathbf{q}) \rangle} = 8\pi^2 \frac{\delta_*(0)}{|\mathbf{q}|^D}.
\end{equation}
The mean square relative displacement $B(\bvec{r}_1-\bvec{r}_2) = \overline{ \langle (u(\bvec{r}_1)-u(\bvec{r}_2))^2 \rangle } $ is calculated as
\begin{eqnarray}
&&B(\bvec{r}_1-\bvec{r}_2) \nonumber \\
&&= 16\pi^2 \delta_*(0) \int \frac{d^D \mathbf{q}}{(2 \pi)^D} \frac{1-\cos \left\{ \mathbf{q} \cdot (\bvec{r}_1-\bvec{r}_2) \right\} }{|\mathbf{q}|^D} \nonumber \\
&&\sim 2 \delta_*(0) {\rm ln} |\bvec{r}_1-\bvec{r}_2|,
\end{eqnarray}
where the prefactor resulting from the surface area of the unit sphere is evaluated at $D=4$.
Therefore, the correlation function is calculated as
\begin{eqnarray}
C(\bvec{r}) = \overline{ \langle e^{i(u(r)-u(0))} \rangle } = \exp \left[ -\frac{1}{2} B(\bvec{r}) \right] \sim |\bvec{r}|^{-\eta},
\end{eqnarray}
with $\eta=\delta_*(0)$.

\subsection{Nonequilibrium case}

We next show Eq.~(\ref{eta_delta_neq}).
The correlation of the phase parameter $u$ is given by
\begin{equation}
\overline{\langle u(\mathbf{q}) u(-\mathbf{q}) \rangle} = \frac{\Delta(0)}{K^2 |\mathbf{q}|^4 + v^2 q_x^2}.
\end{equation}
From Eq.~(\ref{def_delta_neq}), $\Delta(0)$ scales as $\Delta(0)=4\pi K v k^{\epsilon} \delta_*(0)$ at $D=3-\epsilon$, where the prefactor is evaluated at $D=3$.
If the renormalization scale $k$ is replaced with the transverse momentum $|\mathbf{q}_{\perp}|$, we have 
\begin{equation}
\overline{\langle u(\mathbf{q}) u(-\mathbf{q}) \rangle} = 4\pi K v \frac{|\mathbf{q}_{\perp}|^{\epsilon} \delta_*(0)}{K^2 |\mathbf{q}_{\perp}|^4 + v^2 q_x^2},
\end{equation}
where we have omitted the terms containing $q_x$ in $|\mathbf{q}|^4$ because they are negligible compared to $v^2 q_x^2$ in the small momentum regime.
The mean square relative displacement for the transverse direction $(\bvec{r}_1-\bvec{r}_2 \perp \bvec{e}_x)$ is calculated as
\begin{eqnarray}
&&B(\bvec{r}_1- \bvec{r}_2) \nonumber \\
&&= 8\pi K v \delta_*(0) \int \frac{d^D \mathbf{q}}{(2 \pi)^D} |\mathbf{q}_{\perp}|^{\epsilon} \frac{1-\cos \left\{ \mathbf{q} \cdot (\bvec{r}_1-\bvec{r}_2) \right\} }{ K^2 |\mathbf{q}_{\perp}|^4+v^2 q_x^2} \nonumber \\
&&= 4 \delta_*(0) \int \frac{d^{D-1} \mathbf{q}_{\perp}}{(2 \pi)^{D-1}} \frac{1-\cos \left\{ \mathbf{q} \cdot (\bvec{r}_1-\bvec{r}_2) \right\} }{|\mathbf{q}_{\perp}|^{D-1}} \nonumber \\
&&\sim 2 \delta_*(0) {\rm ln} |\bvec{r}_1-\bvec{r}_2|,
\end{eqnarray}
where the prefactor resulting from the surface area of the unit sphere is evaluated at $D=3$.
Therefore, the correlation function is given by $C(\bvec{r}) \sim |\bvec{r}|^{-\eta_{\perp}}$, with $\eta_{\perp}=\delta_*(0)$.

\end{document}